\documentclass[10pt, conference]{IEEEtran}	

\usepackage{bm}
\usepackage{amsmath,amsthm}
\usepackage{subfigure}
\usepackage{footnpag}			      	
\usepackage{cite, graphicx, subfigure,setspace,algorithm}
\usepackage{amsmath,amssymb}
\usepackage{bm}
\usepackage{subfigure}
\usepackage{CJK}

\newtheorem{assumption} {Assumption}
\newtheorem{proposition} {Proposition}
\newtheorem{remark}{Remark}
\newtheorem{theorem}{Theorem}

\usepackage{epstopdf}

\usepackage{graphicx}

\DeclareGraphicsExtensions{.png}

\graphicspath{ {./graphic/} }

\newcommand{\sn}{\sigma^{(n)}}
\newcommand{\sni}{\sigma^{(n)}_i}

\newcommand{\qn}{q^{(n)}}
\newcommand{\rk}{r^{(k)}}
\newcommand{\sib}{\sigma^b}

\newcommand{\qr}{q^{(r)}}
\newcommand{\vn}{v^{(n)}}
\newcommand{\sd}{\sigma^d}
\newcommand{\nq}{\nu^{(q)}}
\newcommand{\rrq}{r^{(q)}}

\usepackage{epstopdf}

\begin{document}

\title{A Belief Space Perspective of RFS based Multi-Target Tracking and its Relationship to MHT}

\author{{S. Chakravorty\thanks{Department of Aerospace Engineering, Texas A\&M University, College Station, TX.},  
W. R. Faber\thanks{Applied Defense Solutions/ L3 Corp., Colorado Springs, MD.},
\ Islam I. Hussein\thanks{Applied Defense Solutions, Columbia, MD},
\ and U. R. Mishra \thanks{Department of Aerospace Engineering, Texas A\&M University, College Station, TX.}
}
}

\maketitle{} 		

\begin{abstract}
 In this paper, we establish a connection between Reid{'}s HOMHT and the modern Random Finite Set (RFS)/ Finite Set Statistics (FISST) based methods for Multi-Target Tracking. We start with an RFS description of the Multi-Target probability density function (MT-pdf), and derive the prediction, and update equations of the MT-tracking problem in the RFS framework from a belief space perspective. We show that the RFS pdf has a hypothesis dependent structure that is similar to the HOMHT hypotheses structure. In particular, we examine the different hypotheses, and derive the hypotheses update equations under the FISST recursions, and clearly show its relationship to the classical HOMHT hypotheses and hypothesis weight update formula, thereby establishing a connection between the methods.
\end{abstract}

\section{Introduction}
\IEEEPARstart{I}{n} this paper, we show that the FISST recursions \cite{Mahler:07} have a discrete hypothesis based structure analogous to the hypothesis oriented multiple hypothesis tracking (HOMHT) \cite{Reid1979}. In order to facilitate this, we derive the FISST based Bayesian Multi- Target probability density function (MT-pdf) recursion for unlabeled target states from a belief/ pdf space viewpoint. 
We show that, in the absence of target birth and death, the RFS hypotheses are identical to the HOMHT hypotheses, and their Bayesian update equations are identical too. For the case with Target birth, the two are different,  nonetheless, under certain simplifying assumptions, the hypotheses update equations for the two become identical given all new births are detected. Moreover, the MHT heuristic of detecting all new births is nonetheless a justifiable approximation that helps keep the problem computationally tractable.

In the last 20 years, the theory of RFS-based multi-target detection and tracking has been developed based on the theory of finite set statistics (FISST) \cite{Mahler:97,Mahler:07}.  The greatest challenge in implementing FISST in real-time, which is critical to any viable multi-target tracking solution, is computational burden. The first-moment approximation of FISST  is known as the Probability Hypothesis Density (PHD) approach \cite{Mahler:07,Vo06}. The PHD has been proposed as a computationally tractable approach to applying FISST. The PHD filter can provide information about the number of targets (integral of the PHD over the region of interest) and likely location of the targets (the peaks of the PHD). The PHD can further employ a Gaussian Mixture (GM) or a particle filter approximation to reduce the computational burden (by removing the need to discretize the state space). This comes at the expense of approximating the general FISST pdf with its first-moment \cite{Vo06, Vo_SMC, Vo_CPHD, Clark_PHD}. The PHD filter does not attempt to solve the full FISST recursions, in particular, by considering the PHD, the filter gets rid of the data association problem inherent in these problems. 
More recently,  the generalized labeled multi-Bernoulli (GLMB) filter has been proposed for the full bayesian MT recursion, that has a hypotheses dependent structure. Tractable implementation of the filter have been proposed based on a lookahead strategy based on the cheaper PHD filter \cite{MB_FISST1, MB_FISST2} as well as based on Gibbs sampling of the hypotheses \cite{MB_FISST3}. The GLMB seeks to label the target states in order to generate "tracks" of targets as in classical MHT. 

There exist non-FISST based ``classical" approaches to multi-target tracking such as the joint probabilistic data association (JPDA) filter, the Hypothesis Oriented MHT (HOMHT) \cite{Reid1979, BarShalom1, BarShalom2, mcmcda}, and the track oriented MHT (TOMHT) techniques \cite{TOMHT}.  These techniques can be divided into single-scan and multi-scan methods depending on whether the method uses data from previous times to distinguish the tracks \cite{BarShalom1, jpda, mcmcda}. The single-scan (recursive) methods such as joint probabilistic data association (JPDA) \cite{jpda, mcmcda} maintain a single hypothesis and typically make the assumption that the tracks are independent which is not necessarily true. The multi-scan methods such as HOMHT and TOMHT \cite{TOMHT, mcmcda} maintain multiple hypotheses at any time step that they disambiguate sequentially using subsequent measurements. 

In this paper,  we show that the full FISST pdf has a hypothesis dependent structure identical to MHT, where the set of hypotheses is the combination of all possible birth, death and data association hypotheses. Albeit the FISST and HOMHT recursions are identical for the case without target birth or death,  the FISST recursions differ from the MHT recursions in the way target birth is handled.
Our development is different from GLMB in that we never explicitly label the targets, in lieu, we show that the target pdfs/ beliefs, under the different hypothesis, are already implicitly uniquely labeled in the FISST recursion. 
The main thesis is that if the problem is studied in the pdf/ belief space, hypotheses and target identity are present in the RFS Bayesian recursions. This is implicit in the HOMHT derivation as well \cite{Reid1979} where the data associations are to pdfs, and there is no explicit labeling involved.
There has been work in recent years, including our own, that seek to relate the structure of MHT to FISST based methods, for instance, the references \cite{Faber_Fusion_2015, Faber_Fusion_2016, RFS-TOMHT,brekke1,brekke2,Mori3f} that draw an analogy between the hypothesis dependent structure of MHT and FISST based multi-target tracking. Our development here differs in that we rely on a belief space perspective, (tracks are beliefs), to study the Bayesian recursions, and a spatial binomial process for the birth model that lets us maintain uniqueness of the birthed pdfs, whether detected or not, but nonetheless lets us connect back to the usual Poisson birth model. We show that the hypotheses and target identity is maintained in RFS based multi-target tracking without the need for explicit labeling.
We derive the hypotheses update formulae inherent in the FISST based method and show the connection to classical HOMHT in terms of these update equations (Eqs. 30 and Eqs. 34-36 in sec IV). In particular, these are derived under different assumptions from the corresponding development in \cite{brekke2}. We also consider the case of undetected birth in detail, and show that the MHT ``heuristic"/ practice of detecting all births by seeding new tracks at the measurements, is a valid approximation of the FISST hypotheses in the sense that any hypothesis with undetected births has an associated hypothesis with no undetected births that always has significantly higher weight. 

\subsection{Outline of the Paper}
The rest of the paper is organized as follows. In Section II, we introduce the FISST MT-pdf and the MT Markov transition and likelihood functions. In section III, we derive the hypothesis dependent structure of the FISST Bayesian recursion starting with the multi-target likelihood and transition functions defined on the target state space. We show that hypothesis and target identity can be maintained in the FISST recursions without explicit labeling when studied in pdf/ belief space. In section IV, we show the relationship between the RFS and HOMHT based approaches to multi-target tracking terms of the hypothesis update equations. We also consider the case of undetected births in this section and show why the MHT ``heuristic" of detecting all births yields a tractable approximation of the hypothesis space. We understand that the notation can seem overwhelming, albeit the underlying concepts are reasonably straightforward, and thus, we have extensively illustrated the different developments with pictures.\\
The application of the FISST to multi-target tracking in Space Situational Awareness (SSA) problems utilizing a randomized approximation of the full FISST recursions, called the RFISST, has appeared in several publications, and is not covered here \cite{Faber_Astro_2015, Faber_Astro_2016, Faber_JAS19, R-FISST}. 

\section{Preliminaries}

In this section, the preliminaries required for the Bayesian MT-pdf recursions are very briefly laid out, the relevant details can be found in \cite{Mahler:07}. 
\subsection{Initialization}
\begin{figure}[hbt]
 \centering
	\includegraphics[width=1.0\linewidth]{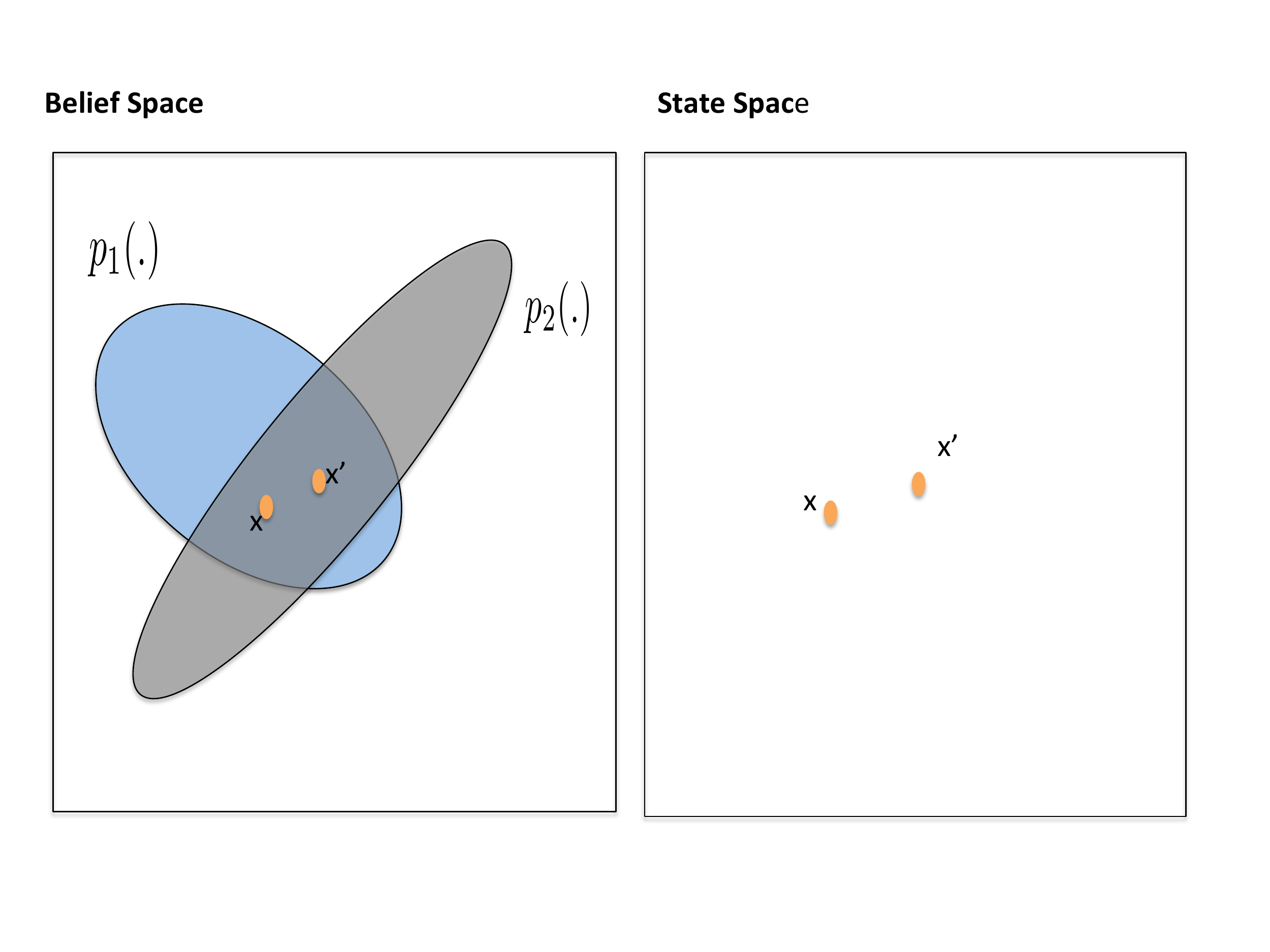}
	\vspace{-0.5in}
	\caption{The initial condition pdf. The pdf is symmetrized, i.e., written as $p(x,x') = p_1(x)p_2(x') + p_1(x')p_2(x)$, because the events E1:  target 1 is at $x$ and target 2 is at $x'$, and E2: target 1 is at $x'$ and target 2 is at $x$, are indistinguishable. This phenomenon shows up on the right in a realization of the process in the state space where it is impossible to say if the sample $x$ came from target/ pdf 1and $x'$ came from target/ pdf 2, or vice-versa. However, in the pdf/ belief space as shown on the left, the pdf of target 1 \textbf{is distinguishable} from that of target 2 (since the pdfs are distinct), i.e., there is no need for ``labeling" them 1 and 2 to distinguish them. The symmetrized pdf representation above is simply a Mathematical representation of the belief space picture on the left.}
	\label{IC}
\end{figure}

Assume, initially, there exist $n$ targets with distinguishable pdfs. The initial distribution can be represented as follows: 
\begin{equation}
\label{initialRFISST}
p_{0}(X,n) = p_{0}(X/n)\rho_0(n),  
\end{equation}
where
$
p_{0}(X/n) = \sum_{\nu}\prod_{i=1}^n p_{0}^i(x_{\nu_i}),
$ 
i.e.,  the multi-target PDF has cardinality $n$, ($\rho_0(n) = 1$), and has independent target states with $p_{0}^i(.)$ denoting the ``distinguishable"  PDF of the $i^{th}$ target. The argument of the pdfs,  $x_{\nu_i}$, \textbf{do not denote the state of the $\nu_i^{th}$ target}, rather they are just points in the target state space. Note that albeit the targets themselves might be indistinguishable in state space resulting in the ``symmetrized" MT-pdf, the pdfs themselves are distinguishable. An illustration of the concept is shown in Fig. \ref{IC}.
\subsection{Multi-target Markov Transition Density Function}
In this section, the multi-target motion model is briefly overviewed for two different cases. 

\textit{Fixed Number Of targets:}
For a general time step, the FISST one step multi-target transition density function, in the absence of birth or death, is given by the following:
\begin{align}
\label{fixedT}
p(X,n| X',n) = \sum_{\nu} \prod_{i=1}^n p(x_{\nu_i}|x'_i),
\end{align}
where $p(x_{\nu_i}|x'_i)$ denotes the corresponding single target transition density function, and $\nu = (\nu_1,\cdots \nu_n)$ represents all possible permutations of the numbers $1$ through $n$.\\
\begin{figure}[hbt]
 \centering
	\includegraphics[width=.9\linewidth]{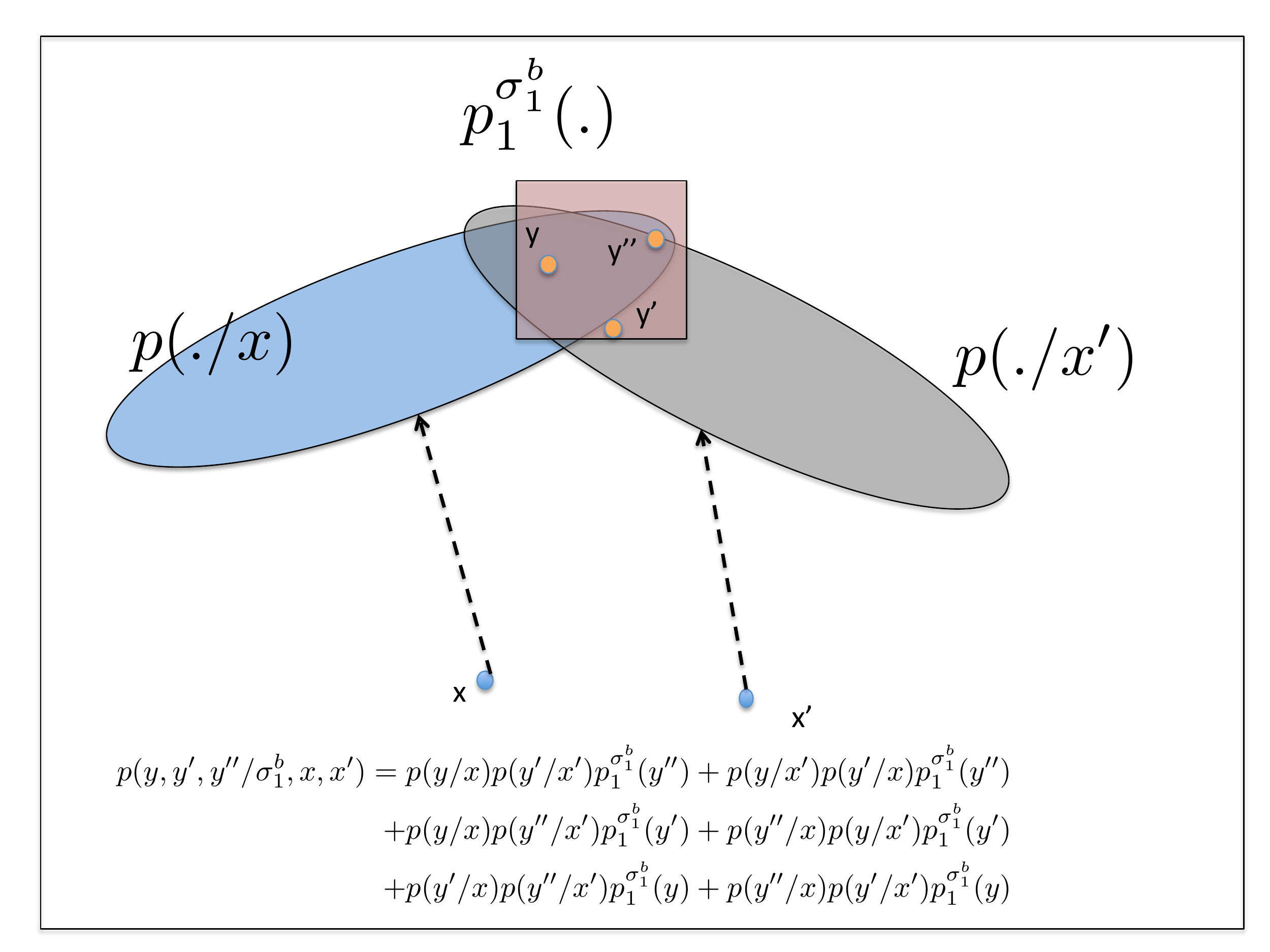}
	\caption{The transition pdf given the birth hypothesis $\sigma^b_1$ that hypothesizes a single birth with pdf $p^{\sigma^b_1}_1(.)$, and also given two targets at $x$ and $x'$ respectively. Albeit the transition pdf looks fearsome, it really only is a result of the symmetrization required of the pdfs due to the indistinguishability of the targets. However, note that in pdf space, the pdfs due to the target at $x$, $x'$ and the birth \textbf{are distinguishable}. The different single birth hypotheses $\sigma^b_1$ correspond to all possible locations a single birth can occur.}
	\label{Binomial}
\end{figure}
\textit{With target Birth:}
In a problem with fixed number of targets, an $r$-target configuration could only result in another $r$-target configuration, however, now due to target birth an $r$-target model can result in an $n$-target model where $n>r$. Assume that one can encode a transition including any arbitrary number of births into a ``birth hypothesis". The multi-target transition function, conditioned on the birth hypotheses and using the Law of Total Probability, is given by:
\begin{align}
p(X,n| X',r) = \sum_{\sib_{n-r}} p(X,n|\sib_{n-r},X',r)p(\sib_{n-r}), \label{birthT}
\end{align}
where  $\sib_{p}$ represents a birth hypothesis that results in exactly $p$ births, $p(\sib_{p})$ is the probability of the birth hypothesis. In the case where a spatial binomial process is assumed for the birth model, $p(\sib_{p})$ is equal to $\alpha^{p}(1-\alpha)^{(M-p)}$ for all p-birth hypothesis $\sib_{p}$. In this probability, $M$ is a measure of the number of possible spatial birth PDF within the sensor field of view (FOV). For example, this can be assumed to be the number of pixels in the sensor FOV, thereby $\sib_p = \{\sib_{p,1}, \sib_{p,2}...\sib_{p,p}\}$ where $\sib_{p,j}$ denotes the pixel corresponding to the $j^{th}$ birth under $\sib_p$. Furthermore, the first factor in the summation of Eq. \ref{birthT} is as follows:
\begin{align}
p(X,n| \sib_{n-r}, X',r) = \sum_{\nu}\prod_{i=1}^r p(x_{\nu_i}| x_i') \prod_{i=r+1}^{n} p^{\sib_{n-r}}_{i-r}(x_{\nu_i}),
\end{align}
where $p^{\sib_p}_j(.)$  denotes the $j^{th}$ birth PDF under the p-birth hypothesis $\sib_{p}$. We assume that the birth pdf is uniformly distributed within the volume of the pixel $\sib_{p,j}$, where $\bar{V}$ denotes the volume of the pixel.
\begin{remark}
We use the binomial birth model to simplify the derivation of the prediction and update equations in Sec. III, in particular, it allows us to distinguish the "pdfs" of the birthed targets in different pixels. We show the connection to the standard Poisson birth model in Section IV: as the number of pixels $M$ becomes large, the pixel volume $\bar{V}$ becomes small, and $\alpha = \lambda \bar{V}$, we approach the Poisson limit for the binomial distribution. In the standard Poisson birth model, the number of births is distributed as a Poisson random variable while the birth pdfs are uniform over the sensor FOV, i.e., indistinguishable from each other. However, these two descriptions are equivalent.
\end{remark}
\begin{figure}[hbt]
 \centering
	\includegraphics[width=.9\linewidth]{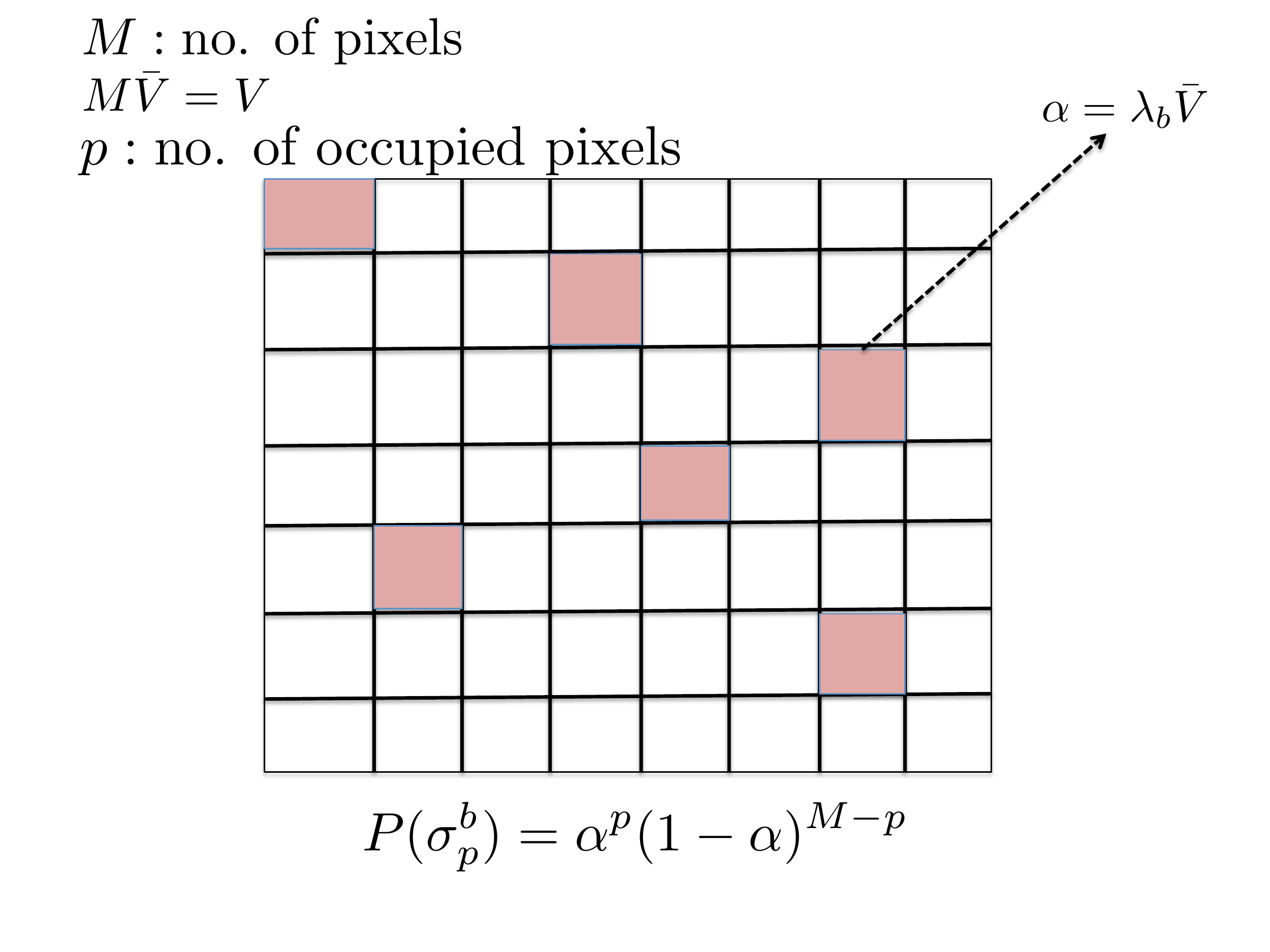}
	\caption{The birth model used in this paper is a spatial Binomial process noting that the Poisson distribution is a limiting case of the Binomial distribution. Each pixel corresponds to a unique birth pdf, and the connection to the Poisson model is obtained by suitable interpretations of the number of pixels $M$, the volume of the pixels $\bar{V}$, and the probability of a pixel being occupied, $\alpha$.}
	\label{Binomial}
\end{figure}
\subsection{Multi-target Likelihood Function}
Given a set observation $\{z_1,z_2\cdots z_m\}$, let $\sn = (\sn_1,\sn_2,..\sn_n)$ denote a data association hypothesis for $n$ targets at the co-ordinates $\{x_1,x_2,\cdots x_n\}$. The association hypothesis, $\sn_i \in \{z_1,z_2\cdots z_m, \phi\}$, associates the $i^{th}$ target to one of $m$ measurements, $z_{\sn_i}$, or to nothing, ${\phi}$. The multi-target likelihood conditioning on all such possible data associations $\sn$ , and using the Law of Total Probability, is given by:
\begin{align}
p(Z|X,n) =  \sum_{\sn}p(Z|\sn, X,n)p(\sn|X,n). \label{MT_lhoodfull}
\end{align}
The term $p(\sn| X,n)$ denotes the a priori probability of the data association $\sn$, given that there are $n$ targets, and that it assigns exactly $k$ measurements to clutter that is Poisson distributed, is given by:
 \begin{equation}
 p(\sn| X,n) = p(\sn| n) 
 = {p_D^{(m-k)}(1-p_D)^{n-(m-k)}}\frac{e^{-\lambda V} (\lambda V)^k}{k! }.\label{aprioriPoisson}
 \end{equation}
Further, the likelihood of the measurement $Z$, given the data association hypothesis $\sn$, is:
\begin{align}
p(Z| \sn, X,n) = \frac{k!}{V^k}\prod_{i=1}^{n} p(z_{\sni}|x_i),\label{fullLikelihoodPoiss} 
\end{align}
assuming that the clutter is uniformly distributed in the sensor volume $V$.\\

\section{A Belief Space Perspective of RFS Bayesian Multiple target Tracking with Fixed Number of targets} \label{hfisst}
In this section, the prediction and update steps for a general multi-target tracking problem are discussed. 


\subsection{Prediction and Update After Initialization for A Fixed Number Of targets}
\begin{assumption} \label{assumpOne}
	Let the multi-target pdf at time $t=0$ have exactly $n$ distinguishable components as defined in Eq. \ref{initialRFISST}:  
	\begin{equation}
	p_{0}(X,n) = \sum_{\nu}\prod_{i=1}^n p_{0}^i(x_{\nu_i}) \rho_0(n),
	\nonumber
	\end{equation} 
	where $\rho_0(n)=1$.
\end{assumption}

\noindent Let the  observation at time $t=1$, be denoted as $Z_1 = \{z_1^1, z_2^1, \cdots z_m^1\}$. Let $p_1^-(X,n)$ denote the predicted MT-pdf just before receiving $Z_1$. Then, the following result holds:
\begin{proposition} \label{Prop1}
	Under Assumption \ref{assumpOne}, the predicted pdf $p_1^-(X,n)$ at time $t=1$ is given by:
	\begin{equation}
	p_1^-(X,n) = p_1^-(X|n) \rho_0(n), 
	\end{equation}
	where
	\begin{equation}
	p_1^-(X|n) = \sum_{\nu} \prod_i p_1^{i-}(x_{\nu_i}), \nonumber\\
	\end{equation}
	\begin{equation}
	p_1^{i-}(x) = \int p(x| x')p_0^i(x') dx',\nonumber
	\end{equation}
	i.e., the predicted multi-target pdf is simply the product of the predicted pdfs of the individual targets. 
\end{proposition}

\begin{proposition}\label{Prop2}
	Under Assumption 1, the updated multi-target PDF at time $t=1$ is given by:
	\begin{align}
	p_1(X,n| Z_1) = \sum_{\sn}p_1^{\sn}(X|Z_1) \omega^{\sn}_1,
	\end{align}
	where $\sn$ is a data association "hypothesis" given that there are $n$ targets and the sum is over all such data associations, and
	\begin{equation}
	p_1^{\sn}(X|Z_1) = \sum_{\nu}\prod_i p_1^i (x_{\nu_i}| z_{\sni}^1),
	\end{equation}
	\begin{equation}
	p_1^i(x| z) = \begin{cases}
	\frac{p(z|x)p_1^{i-}(x)}{\int p(z|x')p_1^{i-}(x')dx'}, ,& \text{if } z \neq \phi\\
	p_1^{i-}(x),              & \text{if} \ z = \phi
	\end{cases} 
	\end{equation}
	\begin{align}
	\omega^{\sn}_1 = \frac{ l_{\sn}p(\sn|n)\rho_0(n)}{\sum_{q, \nq} l_{\nq}p(\nq| q)\rho_0(q)}, \label{E1'}
	\end{align}
	where $p(\sn|n)$ is found from Eq. \ref{aprioriPoisson}, and
	\begin{align}
	l_{\sn} 
	= \frac{k!}{V^k}\underbrace{\prod_i \int p(z_{\sni}^1|x) p_1^{i-}(x)dx}_{\bar{l}_{\sn}}, \label{E2}
	\end{align} 
	assuming that $\sn$ assigns exactly $k$ measurements to clutter. Note that $\sn$ implicitly assumes a sum over all $k$, again this is not shown explicitly purely for notational convenience.
\end{proposition}

\subsection{Update for a General Time Instant}
\begin{figure}[hbt]
 \centering
	\includegraphics[width=1.0\linewidth]{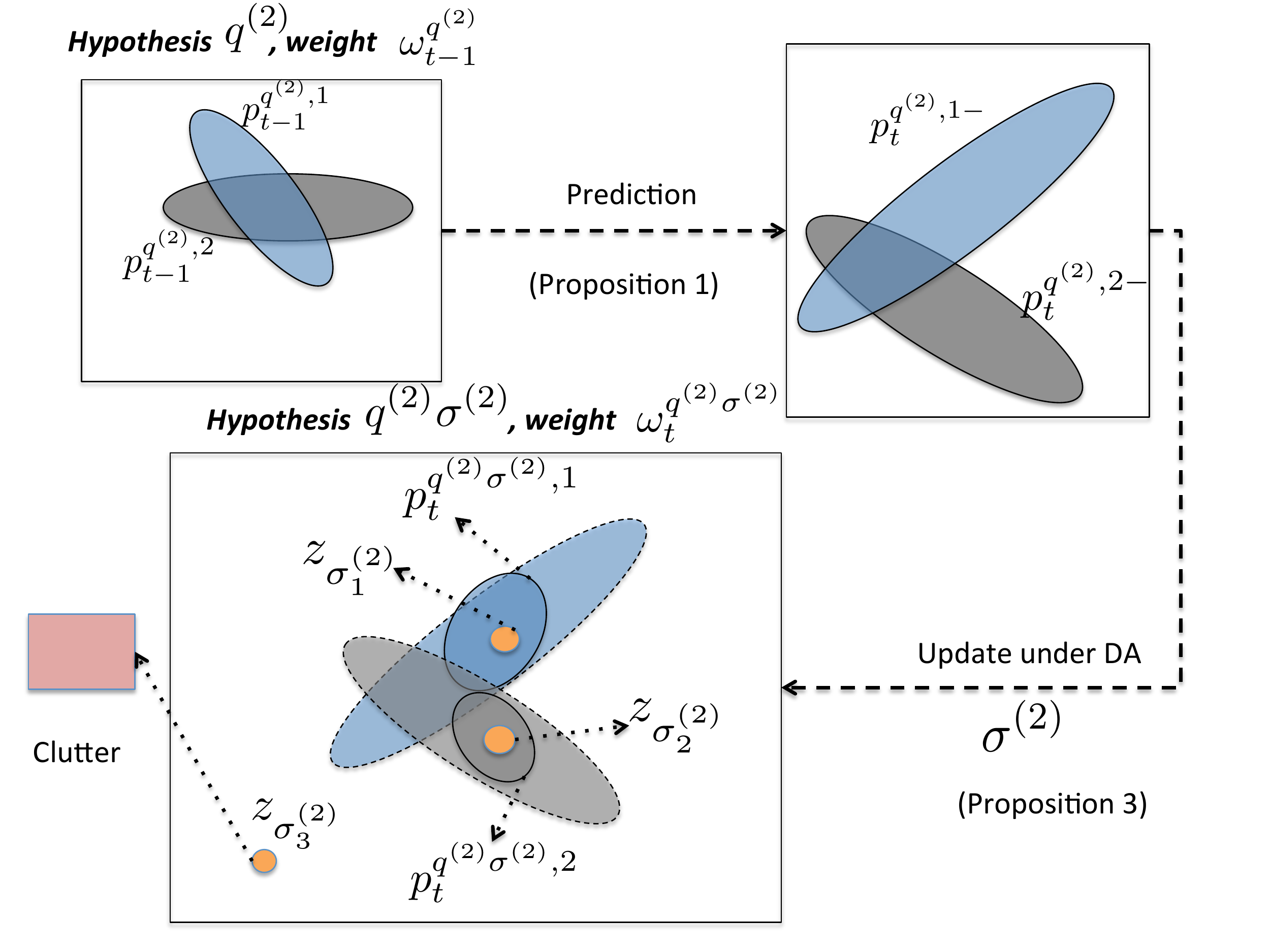}
	\caption{An Illustration of Propositions 1-3 showing Bayesian prediction and update under FISST. The component pdfs 1 and 2 at time $t-1$ under hypothesis $q^{(2)}$ are predicted to obtain the predicted prior at time $t$. Given the data association $\sigma^{(2)}$, the pdfs 1 and 2 are updated according to the measurements $z_{\sigma^{(2)}_1}$ and $z_{\sigma^{(2)}_2}$ respectively while $z_{\sigma^{(2)}_3}$ is associated to clutter. The fearsome looks of the FISST pdfs (in Eq. \ref{M_c}) is again due to the symmetrization necessary to account for the \textit{indistinguishability of target states}. Moreover, the \textbf{data associations are to pdfs which are distinguishable and not to the ``labels" 1,2; the labels are purely incidental and can be swapped without changing the MT-pdf. This is further elaborated in Sec III C.}}
	\label{FISST_Bayes}
\end{figure}

Let the general multi-target PDF at time $t-1$ can be represented by,
\begin{equation}
\label{generalQPDF}
p_{t-1}(X,n)  = \sum_{\qn} p_{t-1}^{\qn}(X) \omega^{\qn}_{t-1},
\end{equation}  
Where the sum is taken over all possible parent hypotheses $\qn$ containing $n$ targets and $\omega^{\qn}_{t-1}$ is the corresponding weight.
The first factor of \ref{generalQPDF} corresponds to the underlying multi-target state given the particular set of $n$ targets, $\qn$, and is expressed,
\begin{equation}
p_{t-1}^{\qn} (X) = \sum_{\nu}\prod_i p_{t-1}^{\qn, i} (x_{\nu_i}). 
\end{equation} 
The predicted PDF at time $t$ is given by:
\begin{equation}
\label{fixedPrediction}
p_t^-(X,n) = \sum_{\qn} p_t^{\qn -}(X) \omega_{\qn},
\end{equation}
\begin{equation}	
p_t^{\qn -}(X) = \sum_{\nu}\prod_i p_t^{\qn, i -}(x_{\nu_i}) ,\nonumber\\
\end{equation}
\begin{equation}
p_t^{\qn, i -}(x) = \int p(x|x') p_{t-1}^{\qn, i}(x') dx'.\nonumber\\
\end{equation}
\begin{proposition}\label{Prop3}
	Further, given a measurement $Z_t$, the updated multi-target PDF is given by:
	\begin{equation}
	\label{fixedUpdate}
	p_t(X,n| Z_t) = \sum_{\qn} \sum_{\sn} p_t^{\qn\sn} (X|Z_t) \omega^{\qn\sn}_t, 
	\end{equation}
	where 
	\begin{align}
	\omega^{\qn\sn}_t = \frac{\omega^{\qn}_{t-1}p(\sn|n)l_{\qn\sn}}{\sum_q \sum_{\rrq}\sum_{\nq} \omega_{\rrq}p(\nq|q) l_{\rrq\nq}}, \label{M_d}\\
	l_{\qn\sn} 
	= \frac{k!}{V^k} \prod_i \int p(z_{\sni}^t|x) p^{\qn, i -}_t (x)dx,
	\end{align}
	\begin{align}
	p_t^{\qn\sn} (X|Z_t) \equiv \sum_{\nu} \prod_i p^{\qn\sn, i}_t(x_{\nu_i}|z^t_{\sni}), \nonumber\\
	p^{\qn\sn, i}_t(x|z^t_{\sni}) = \frac{p(z^t_{\sni}|x)p^{\qn , i -}_t(x)}{\int p(z^t_{\sni}|x')p^{\qn , i -}_t(x') dx'} \label{M_c}
	\end{align}
\end{proposition}
An illustration of the results above is shown in Fig. \ref{FISST_Bayes}.\\
\begin{figure}[hbt]
 \centering
	\includegraphics[width=1.2\linewidth]{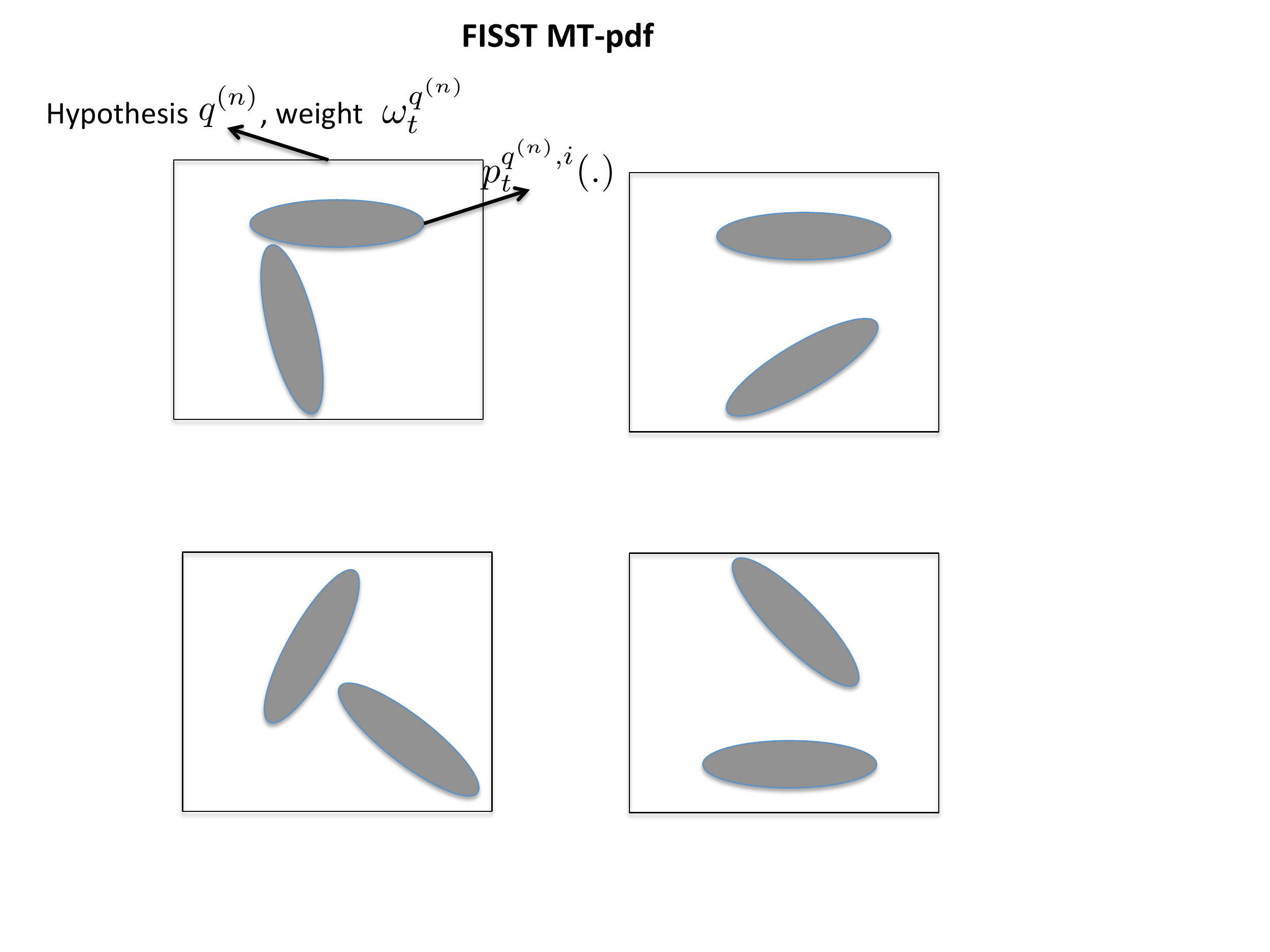}
	\caption{The FISST pdf is a collection of hypotheses $\qn$ with underlying component pdfs $p^{\qn,i}_t(.)$ and weight $\omega^{\qn}_t$. The pertinent question is if these hypotheses can be identified, and if the identity of the targets underlying any hypothesis can be ascertained? The answer, somewhat surprisingly, is YES! }
	\label{FISST-structure}
\end{figure}

\subsection{Hypotheses and Target Identity in FISST based MT-Tracking}
In general, the FISST pdf (from Proposition \ref{Prop3}) at time $t$ is given by the following expression:
\begin{align}
p(X,n) = \sum_{\qn} \omega^{\qn}_t p^{\qn}_t(X), \nonumber\\
p^{\qn}_t(X) = \sum_{\nu} \prod_i p^{\qn,i}_t(x_{\nu_i}), \label{FISSTpdfEq}
\end{align}
where the $\qn$ are the component $n-$ target hypotheses and the $p^{\qn,i}_t(.)$ are the component pdfs underlying the hypothesis $\qn$. An illustration of this is shown in Fig. \ref{FISST-structure}. However, the index $i$ is purely incidental, and does not mean it is the pdf of the $i^{th}$ target.  \\

\textit{This fact raises a question about these hypotheses: can they be identified uniquely? Moreover, can FISST maintain target identity? At first blush, the answer seems negative. However, the answer turns out to be affirmative. We show below how this is indeed the case.}\\

First, let us define a key structure. Let $p_t^{(k,l)}$ denote the $l^{th}$ track of the $k^{th}$ target at time $t$, given by the pdf of the $k^{th}$ target under the measurement sequence $Z^{t,(k,l)} = \{z_{\tau}^{(k,l)}, \tau = 0\cdots t\}$. Here, $z_{\tau}^{(k,l)}$ represents the particular measurement, which could be the null measurement $\phi$,  assigned to target $k$ under the track $l$ at time $\tau$. These tracks are entirely identical to the tracks in the TOMHT formulation \cite{TOMHT}.  \\

Next, by construction (Propositions \ref{Prop1}-\ref{Prop3}), given any hypotheses $\qn$, and an underlying component pdf $p^{\qn,i}_t(.)$, it follows that $p^{\qn,i}_t(.) = p_t^{(k,l)}$ for some unique $(k,l)$, i.e., the component pdf $p^{\qn,i}_t(.)$ corresponds to a unique track of a unique target. Note that the pdf at the track end point is a sufficient statistic/ belief state for the entire history and can be used interchangeably with the history, as is well known from the Partially Observed Markov Decision Problem (POMDP) literature \cite{bertsekas1,kumar1}.\\ 

The next question is: given some hypothesis $\qn$, and some underlying component pdf $p^{\qn,i}_t(.)$ , is it possible to identify the unique track $(k,l)$ corrsponding to it? The answer turns out to be yes, the reason is as follows.\\
The tracks of any target are ``almost surely unique" in the following sense: given a track $(k,l)$, and a different track ${(m,n)}$, in that at least one of the relationships $k\neq m$ and $l \neq n$ holds true, the pdfs corresponding to the tracks are almost surely different, $p_t^{(k,l)} \neq p_t^{(m,n)}$. This holds if the single target transition measure and the single target measurement likelihood measure admit densities as was already assumed in Section II. The proof of this result is presented in the appendix.\\
\begin{figure}[hbt]
 \centering
	\includegraphics[width=0.8\linewidth]{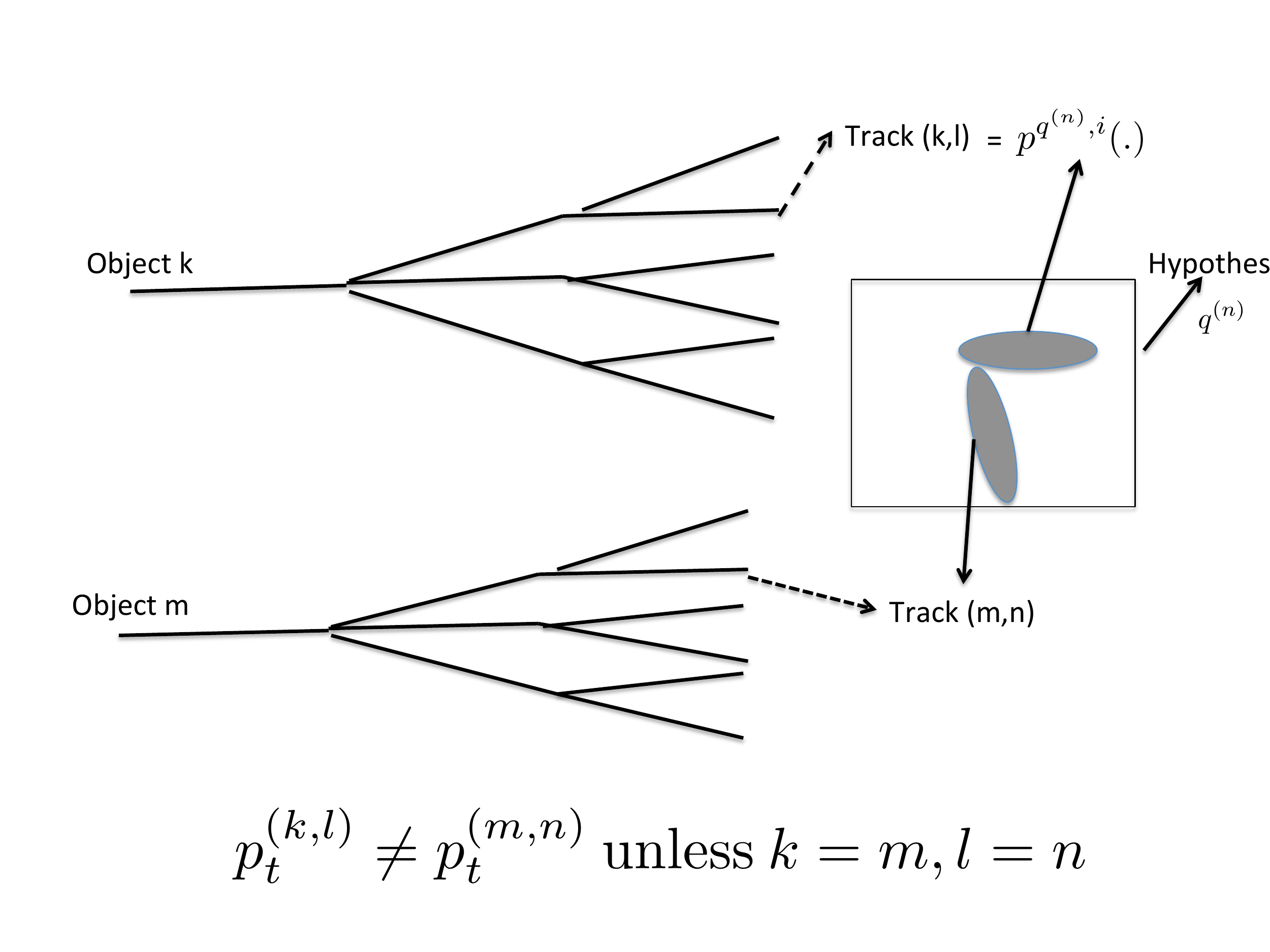}
	\caption{Every track is almost surely distinct from any other track and the track end points are sufficient statistics/ belief states encapsulating the entire history. By construction (Proposition \ref{Prop1}-\ref{Prop3}), the FISST component pdf $p^{\qn,i}_t(.)$ corresponds to a unique track. Since tracks are almost surely distinct, the track end point/ belief state also identifies the track.}
	\label{tracks}
\end{figure}
Thus, we can identify the unique track $(k,l)$ corresponding to the component pdfs $p^{\qn,i}_t(.)$, for any $i$. Therefore, taken together, by construction (Propositions \ref{Prop1}-\ref{Prop3}),  these tracks define the unique hypothesis corresponding to $\qn$. Moreover, the weight of the hypothesis is given by $\omega^{\qn}_t$, again, by construction. Hence, we may summarize the above result as follows.
\begin{theorem}
\textbf{Hypotheses and Target Identity in FISST.} Given the FISST pdf in Eq. \ref{FISSTpdfEq}, the different hypotheses $\qn$, and the underlying tracks $p^{\qn,i}_t(.)$ can be uniquely identified, almost surely.
\end{theorem}
\begin{remark}
The above result establishes the fact that even in FISST, track identity, and consequently, hypothesis identity is assertable, owing to the almost sure uniqueness of the different tracks,  However, this is a purely theoretical construct in that it would require one to keep a ``look up table" of all possible tracks to identify the different  tracks underlying different hypotheses. In practice, however, this clumsy bookkeeping is unnecessary if we simply propagate a unique label corresponding to the different tracks of each target, whenever updating its pdf under a hypothesis. However, this is necessary purely for implementational convenience, and should not be construed as the necessity of labeling. 
\end{remark}

\begin{remark}
The hypothesis/ track based structure of the FISST pdf flows entirely from using the MT-likelihoods and MT-transition functions as specified by FISST, along with Assumption 1 of distinct initial component pdfs (Propositions \ref{Prop1}-\ref{Prop3}). This is different from the corresponding development in \cite{brekke2} that uses a clever definition of ``association variables" and ``association hypotheses". 
\end{remark}

\section{RFS Multi target Tracking With Birth}\label{sec:withBandD}
\begin{figure}[hbt]
 \centering
	\includegraphics[width=1.0\linewidth]{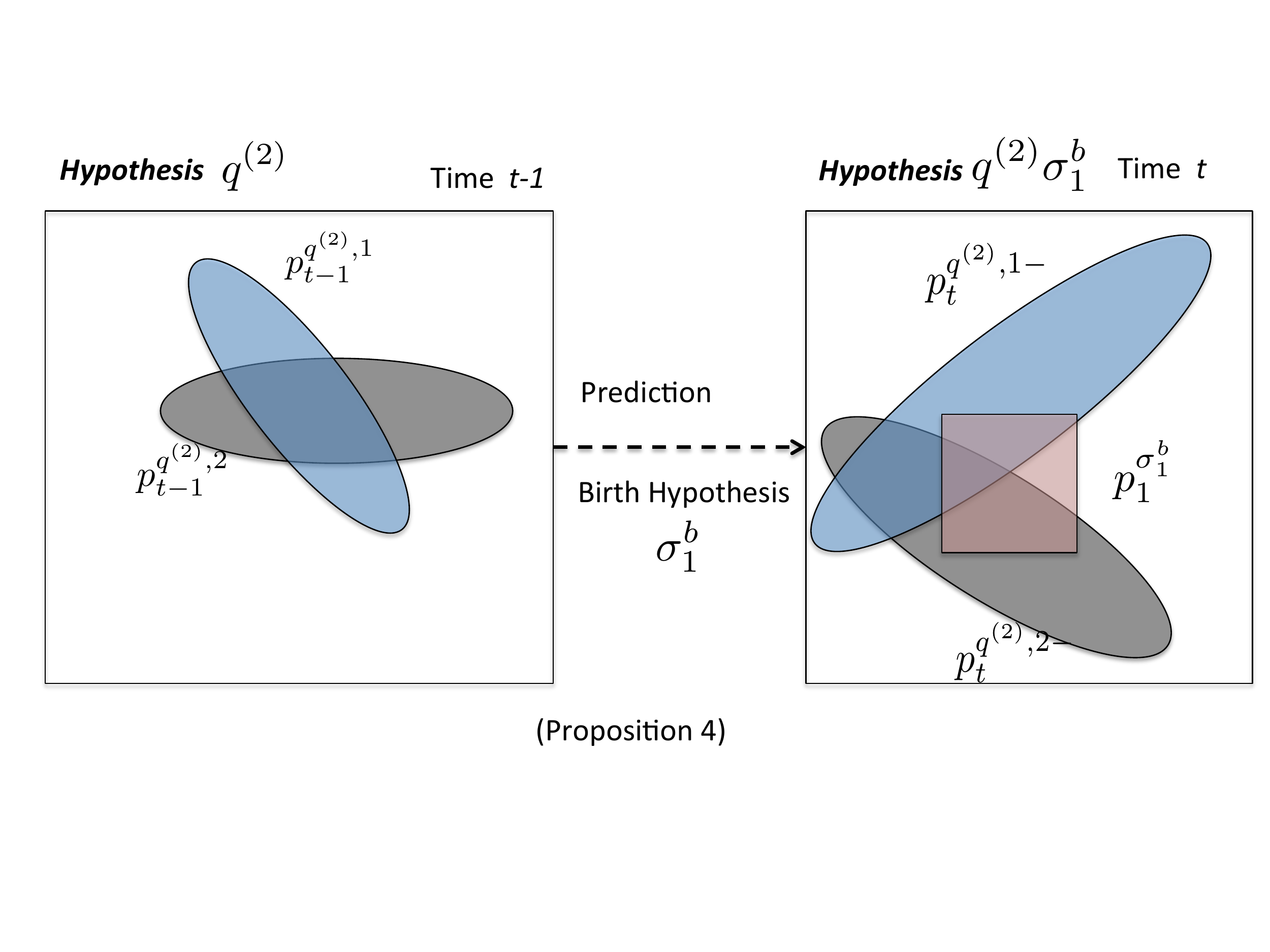}
	\vspace{-0.5in}
	\caption{An Illustration of Propositions 4 showing Bayesian prediction with target birth under FISST. The pdfs 1 and 2 are predicted forward as before. However, due to the single birth hypothesis $\sigma^{b}$, there is an additional birth pdf $p^{\sigma^b_1}_1$ as shown. Again, the labels 1 and 2 are superfluous as the pdfs are distinct.}
	\label{FISST_PB}
\end{figure}

The following proposition describes the prediction and update steps for a multi-target system with a varying number of targets caused by birth and death and includes a measurement model with both missed detections and false alarms.
\begin{assumption}\label{fullPred}
Assume that the multi-target PDF $p(X,r)$ has the structure:
\begin{equation}
p_{t-1}(X,r) = \sum_{\qr} \omega^{\qr}_{t-1}p^{\qr}_t(X),
\end{equation}
\begin{equation}
p^{\qr}_{t-1}(X) = \sum_{\nu}\prod_{i=1}^r p^{\qr ,i}_{t-1}(x_{\nu_i}).
\end{equation}
\end{assumption}
The predicted multi-target PDF under the above birth model is then given by (the proof is provided in the Appendix):
\begin{equation}
p_t^-(X,n) = \sum_r \sum_{\sib_{n-r}} p(\sib_{n-r}) \sum_{\qr} \omega^{\qr}_{t-1} p^{\qr -}_{\sib_{n-r},t}(X), \label{B-P}
\end{equation}
where
\begin{equation}
p^{\qr -}_{\sib_{n-r},t}(X) 
= \sum_{\nu} \prod_{i=1}^r p^{\qr, i-}_t(x_{\nu_i}) \prod_{i=r+1}^{n} p^{\sib_{n-r}}_{i-r}(x_{\nu_i}),\label{P-B}
\end{equation}
\begin{equation}
p^{\qr, i-}_t (x) = \int p(x|x')p^{\qr,i}_{t-1}(x')dx', 
\end{equation}
i.e., the multi-target PDF  is simply the product of the $r$-target predicted PDF given the initial pdf is $p^{\qr}_t(.)$ and the birth hypothesis encoded in $\sib_{n-r}$. It is clear from Eq. \ref{B-P} that the following holds.\\
\begin{proposition} \label{Prop4}
The predicted multi-target PDF, given that the prior multi-target PDF satisfies Assumption \ref{fullPred}, may be expressed as:
\begin{align}
p_t^-(X,n) = \sum_{\vn} \omega^{\vn}_tp^{\vn}_t(X), \label{HFISST-P}
\end{align}
where $\vn = (\qr, \sib_{n-r})$, for all feasible $r$, i.e., each $\vn$ is a combination of some prior $r$-target hypothesis $\qr$ and a corresponding birth hypothesis $\sib_{n-r}$ with 
\begin{equation} \label{weightUpdateWB}
\omega^{\vn}_t = p(\sib_{n-r})\omega^{\qr}_t,
\end{equation}
\begin{equation}
p^{\vn}_t(X) =  p^{\qr -}_{\sib_{n-r},t}(X).\nonumber\\
\end{equation}
where $p^{\qr -}_{\sib_{n-r},t}(X)$ is given by Eq. \ref{P-B}.\\
\end{proposition}	
Given the predicted PDF $p^-(X,n)$ has the hypothesis based form above, with track independence inherent to the multi-target PDF underlying every hypothesis, it is clear that Proposition \ref{Prop3} can be used to perform the update step given a measurement $Z_t$. An illustration of Proposition \ref{Prop4} is given in Fig. \ref{FISST_PB}.

\textit{Hypothesis Weight Update Equation.} 
Let ${\qr}$ denote an initial $r$-target hypothesis, $\sib_{n-r}$ denote a subsequent birth hypothesis, and $\sn_a$ denote
a subsequent data association hypothesis. From Propositions \ref{Prop3} and \ref{Prop4}, it follows that the weight of the grandchild hypothesis (birth followed by data association) of $\qr$ is given by the weight update equation:
\begin{align}
\omega_{\qr\sib_{n-r}\sn_a} = \eta l_{\qr\sib_{n-r}, \sn_a} p(\sn_a|n)p(\sib_{n-r})\omega_{\qr}, \label{D-FISST*}
\end{align}
where $\eta$ is a suitable normalization constant found by summing the numerator over all possible grandchild hypotheses. In the above equation, $p(\sib_{n-r})$ represents the probability of the birth hypothesis, $p(\sn_a|n)$ represents the a priori probability of the data association hypothesis, and $ l_{\qr\sib_{n-r}, \sn_a}$ represents the likelihood of the data association hypothesis $\sn_a$ under the birth hypothesis $\sib_{n-r}$, $\vn = (\qr,\sib_{n-r})$. The predicted MT-pdf underlying $\vn=(\qr,\sib_{n-r})$ is specified by Proposition \ref{Prop4}, and the likelihood is specified by Proposition \ref{Prop3}, given the predicted multi-target PDF. In the next section, a more accessible form of the update will be derived in terms of the parameters of the MTT problem.

\subsection{Hypothesis and Target Identity with Birth}
The next question that we need to answer is whether target tracks and hypothesis identity can be ascertained in the case with birth as was done in the case with a fixed number of targets.\\
After the update step, the FISST pdf with target birth can be written as in Eq. \ref{FISSTpdfEq}. Thus, we need to show that the different component  pdfs $p^{\qn,i}_t(.)$ underlying any hypothesis $\qn$ can be identified with a unique track, and consequently, the hypothesis $\qn$ can be identified. The difference with the fixed number of targets case is that in this case, the tracks can now correspond to any of the birthed pdfs till the current time. Recall that there are $M$ distinct birth pdfs at any time instant, corresponding to the different pixels in which they are born (see Fig. \ref{Binomial}). Thus, all the births till the current time are distinct, and one can keep tracks for each unique birth pixel $\xi$, at every time $\tau$, till the current time time $t$, say denoted by the 2-tuple $(\tau,\xi)$. Again, since the number of distinct birth pdfs is finite in any finite period of time, the tracks of these birthed pdfs are almost surely distinct from any other track, whether from an initial target or any other birthed target. Therefore, analogous to the fixed number of targets case, the unique track $(k,l)$, corresponding to the component pdf $p^{\qn,i}_t(.)$ can be uniquely determined for all $i$, and consequently, the hypothesis $\qn$ can be uniquely identified, while its weight is available by construction.

\begin{remark}
Due to our Binomial birth model, births (in distinct pixels) at the same time can be distinguished, even when undetected. This is different from the standard Poisson birth model where undetected births at the same time would be indistinguishable, nonetheless, the Binomial model can be related back to the standard Poisson model by letting the number of pixels become very large as we shall show below.
\end{remark}

\section{Relationship Between Classical (HOMHT) and RFS based Multi-target Tracking Techniques} 
In this section, the RFS hypotheses update equations from Section III are related to the MHT (specifically HOMHT) hypotheses update equations in order to draw conclusive evidence about the relationship between the methods. There are three parts to this Section:  the first section discusses the relationship between the approaches when there is a fixed number of targets while the following section expands the discussion to include target birth. In the third section, we consider the case of undetected births in the FISST hypotheses.

\subsection{Relationship when the Number of Targets is Fixed} 
Consider a prior n-target hypothesis $\qn$, and assume no birth or death of targets, and the standard measurement model including missed detections and Poisson false alarms. Further, let the predicted densities of the targets underlying $\qn$ be given by $p^{\qn,i-}(.)$. The MHT posterior hypothesis probability, assuming that the data association $\sn$ assigns exactly $k$ measurements to clutter,  is given by \cite[pg. ~847 Eq. ~8]{Reid1979}:    
\begin{align}
\omega_{MHT}^{\qn,\sn} = \eta p_D^{m-k} (1-p_D)^{n-(m-k)}\lambda^k \nonumber\\
 \prod_{i} \int p(z_{\sn_{i}}|x_i)p^{\qn, i-}(x_i)dx_i \; \omega_{\qn}. 
\end{align}
This expression for the MHT posterior probability is exactly the same as the updated weight $\omega^{\qn,\sn}$ presented in Proposition \ref{Prop3} in Eq. \ref{M_d}. Moreover, the updated pdfs of the targets in MHT is identical to the FISST continuous update Eq. \ref{M_c}.\\
The above assertion becomes clear by noting that the factor $\int p(z_{\sn_{i}}|x)p^{\qn, i-}(x)dx$, under a Gaussian approximation, reduces to substituting $z_{\sn_{i}}$ into the Gaussian pdf of the measurement innovation,  $\mathcal{N}(\zeta -H\hat{x}^{\qn,i-}; 0, HP^{\qn,i-}H' + R)$, where the measurement equation has the linear form $z = Hx +v$, the measurement noise $v$ has covariance $R$, and the $i^{th}$ predicted pdf under hypothesis $\qn$, $p^{\qn,i-}$, is assumed to be Gaussian with mean and covariance $\hat{x}^{\qn,i-}$ and $P^{\qn,i-}$: the case considered in Reid's paper. The fact that the target pdf update in MHT is identical to the FISST update Eq. \ref{M_c}, follows from noting that Eq. \ref{M_c} reduces to the Kalman filter if we assume that the predicted pdfs' are Gaussian. Thus, the FISST update can also be construed as a generalization of Reid's update to cases when the linear Gaussian approximation does not hold. 
\subsection{Relationship between HOMHT and FISST with Target Birth}
Consider the weight update equation with birth, Eq. \ref{D-FISST*}.The following development will show how this equation relates to that of \cite{Reid1979}.  Let $q^{(r)}$ denote an $r$-target prior hypothesis. Assuming a binomial process for birth, consider a particular birth hypothesis $\sigma^b_{n-r}$ with associated $n-r$ births at the specified pixel locations $(\sigma^b_{n-r,1}, \cdots , \sigma^b_{n-r,n-r})$. Each spatial density is unique with pdf $p^{\sigma^b_{n-r}}_i(x)= \frac{1_{\sigma^b_{n-r,i}}(x)}{\bar{V}}$ where $\bar{V}$ is the volume of the pixel, and $1_{\sigma^b_{n-r,i}}(x)$ denotes the indicator function on the pixel denoted by $\sigma^b_{n-r,i}$. Furthermore, consider a data association hypothesis $\sn_a$ that associates exactly $s$ of these birth pdf to measurements, and $k$ to clutter. Then, let $\sigma^{(n)}_{a,i}$, $i =1,...,r$ denote the association to existing targets, and for $i =r+1,...,n$ denote the associations to birth. Using Eq. \ref{D-FISST*}, we obtain:
\begin{align}
\omega^{q^{(r)}\sib_{n-r}\sn_a} = \nonumber\\
\eta \frac{k!}{V^k}(\prod_{i=1}^{r}\int p(z_{\sigma^{(n)}_{a,i}}|x)p^{q(r),i-}(x)dx) \nonumber\\
(\prod_{i=r+1}^{n} \int p(z_{\sigma^{(n)}_{a,i}}|x) \frac{1_{\sib_{n-r, i-r}}(x)}{\bar{V}}dx) \nonumber\\
\underbrace{\alpha^{n-r}(1-\alpha)^{M-(n-r)}}_{p(\sigma^b_{n-r})}\nonumber\\
\underbrace{(1-p_D)^{n-(m-k)}p_D^{m-k}e^{-\lambda_CV}\frac{(\lambda_CV)^k}{k!}}_{p(\sn_a | n)}\omega_{\qr}\label{weightCompWBSimp}
\end{align} 
\textit{Poisson Approximation to the Binomial distribution:} Recall the spatial birth model in Fig. \ref{Binomial}. Then, the Poisson approximation to the Binomial distribution states that as $M\rightarrow\infty$, $\bar{V} \rightarrow 0$, with $\alpha = \lambda_B \bar{V}$ and $M\bar{V}=V$,
	\begin{equation}
	\binom{M}{n-r}\alpha^{n-r}(1-\alpha)^{M-(n-r)} = \frac{e^{-\lambda_BV}(\lambda_BV)^{(n-r)}}{(n-r)!}.
	\end{equation}
	The above can be used to yield the following approximation to the birth probability:
\begin{equation}\label{pT}
\alpha^{(n-r)}(1-\alpha)^{M-(n-r)} \approx e^{-\lambda_BV}(\lambda_B)^{(n-r)}\bar{V}^{(n-r)}.
\end{equation}


\noindent
\textit{Assumptions on Measurement Model.} If we assume that $z = x + v$, i.e., a full state measurement, the integral 
$
\int p(z_{\sigma^{(n)}_{a,i}}|x) dx = 1,  
$
if $z_{\sigma^{(n)}_{a,i}}\neq \phi$. Suppose further that the support of the likelihood $p(z/x)$, given $z$, is always within exactly one of the pixels in the spatial birth model. The likelihood confined to one pixel assumption is valid if the support of the measurement likelihood is much smaller than the pixel volume $\bar{V}$, which is a reasonable approximation since the pixel volume is typically large.\\

\noindent
Now, consider the factor:
$
\int p(z_{\sigma^{(n)}_{a,i}}|x). 
\frac{1_{\sib_{n-r,i}}(x)}{\bar{V}}dx.
$
Using the assumptions on the measurement model,
$
\int p(z_{\sigma^{(n)}_{a,i}}|x) 
\frac{1_{\sib_{n-r,i}(x)}}{\bar{V}}dx =  \frac{1}{\bar{V}},
$
if $z_{\sigma^{(n)}_{a,i}}\neq \phi$, or $1$ otherwise, given the support of $p(z_{\sn_{a,i}}|x)$ is within the volume of the pixel $\sib_{n-r,i}$ for all $i$. If even one of the measurements' support is not within one of the pixels specified by the birth hypothesis, then the corresponding weight of the birth hypothesis is zero.\\

\noindent
Substituting the above simplifications into Eq. \ref{weightCompWBSimp},
\begin{align}
\omega_{FISST}^{q^{(r)}\sib_{n-r}\sn_a} = \eta (1-p_D)^{n-(m-k)}p_D^{(m-k)}\nonumber\\
\prod_{i=1}^{r}\int p(z_{\sigma^{(n)}_{a,i}}|x)p^{q(r),i-}(x)dx_i \nonumber\\
(\lambda_B)^{(n-r)}\bar{V}^{(n-r)-s}(\lambda_C)^k\omega_{\qr}.
\label{weightCompSubSimp}
\end{align}
Suppose now that underlying an HOMHT hypothesis, there were $r$ existing targets in the prior hypothesis $\qr$ with the same component pdfs as in the FISST hypothesis above, $n-r$ births were hypothesized,  and exactly $k$ measurements were associated to clutter under the data association $\sn_a$ as in the FISST hypotheses above.  It can be seen from \cite[pg. 848 Eq. 16]{Reid1979} that,  the HOMHT posterior hypothesis probability is, 
\begin{align}
\omega_{HOMHT}^{\qr\sib_{n-r}\sn_a}  = \eta (1-p_D)^{n-(m-k)}p_D^{(m-k)}\nonumber\\
(\prod_{i=1}^{r}\int p(z_{\sigma^{(n)}_{a,i}}|x)p^{q(r),i-}(x)dx)\ (\frac{\lambda_B}{p_D})^{n-r}(\lambda_C)^k\omega_{\qr},\label{weightCompHOMHT}
\end{align}
where it should be noted that \textit{under the HOMHT update, all $n-r$ births are detected, i.e, there is no provision in Reid's HOMHT update for detecting a subset of the births ($ s= n-r$ always).} In general, this is true of MHT techniques in practice as well where new tracks are originated from every measurement \cite{Blackman_MHT}. 
The equations \ref{weightCompSubSimp} and \ref{weightCompHOMHT} clearly show that the FISST approach treats target birth as part of the prediction, and thus, there may be undetected births. 
Suppose now that in the FISST update Eq. \ref{weightCompSubSimp}, we set $s = n-r$, corresponding to the hypothesis that all hypothesized new births are detected. Then Eq. \ref{weightCompSubSimp} becomes,
\begin{align}
\omega_{FISST}^{\qr\sib_{n-r}\sn_a} = \eta (1-p_D)^{n-(m-k)}p_D^{(m-k)}\nonumber\\
(\prod_{i=1}^{r}\int p(z_{\sigma^{(n)}_{a,i}}|x)p^{q(r),i-}(x)dx) 
(\lambda_B)^{n-r}(\lambda_C)^k\omega_{\qr}.
\label{weightCompAllDet}
\end{align}
\begin{figure}[hbt]
	\centering
	\includegraphics[width=1.0\linewidth]{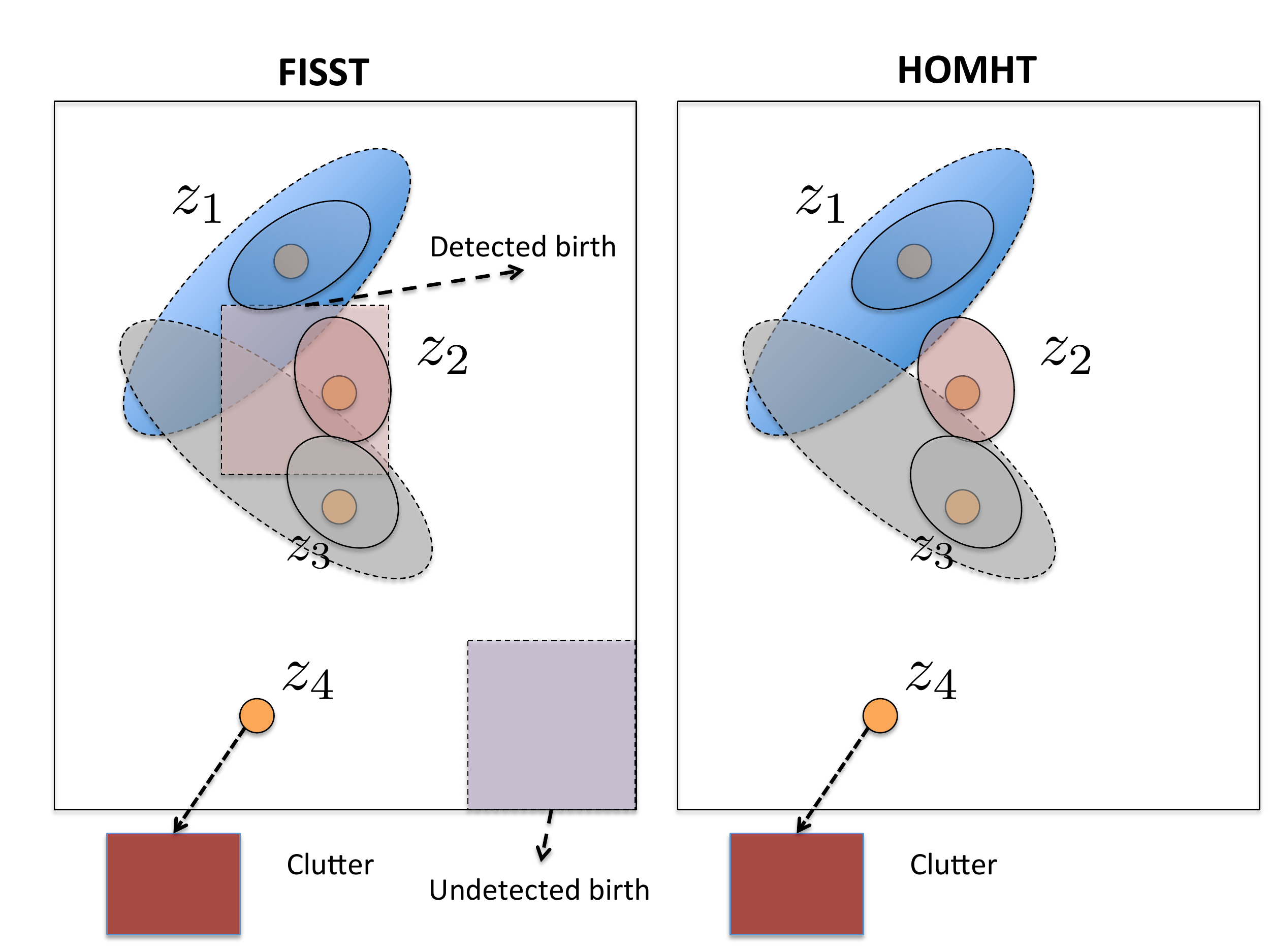}
	\caption{Treatment of Birth in FISST vs HOMHT. The observations $z_1$ and $z_3$ are associated to existing targets in both methods, while $z_4$ is associated to clutter. The observation $z_2$ is associated to the new birth in FISST while a new birth is seeded with the observation $z_2$ in HOMHT leading to the lack of a an extra $p_D$ factor in HOMHT. There is an undetected birth in FISST which is not possible in the HOMHT.}
	\label{MHT_FISST_birth}
\end{figure}
\noindent\textit{Treatment of Birth in FISST vs MHT:} In the case of a hypothesis with no undetected births,  the difference between Eq. \ref{weightCompHOMHT} and Eq. \ref{weightCompAllDet} is the factor  $(\frac{\lambda_B}{p_D})^{n-r}$. Since the HOMHT approach handles birth in the update step the factor is different from that in the FISST based approach, in particular, the HOMHT treats the births the same as clutter but with a different arrival rate. Moreover, the new births are not detected like a standard target, which shows up in the extra $(\frac{1}{p_D})^{n-r}$ factor in the HOMHT recursion. This discrepancy stems from the somewhat nebulous treatment of birth in Reid's original paper where he has both confirmed and ``tentative" targets. However, its not clear from his development how the tentative targets are initialized, and thus, in our interpretation above, we have assumed that there are no ``tentative targets". Later MHT work does include the extra $p_D$ factors and the above discrepancy does not arise in these cases \cite{MHT_review_18}.
However, in general, in FISST, there can be more births than detected whereas there is no such provision in the MHT (albeit there is some work that addresses this issue of undetected births in the MHT framework \cite{MHT-UB}). The differences in the treatment of birth is illustrated in Fig. \ref{MHT_FISST_birth}. 

\subsection{The Case of Undetected Births} 
Let us now take a closer look at the case of undetected births. In the following, we shall show that for any hypothesis that has undetected births, there is an associated hypothesis with no undetected births, whose weight starts higher and always remains higher, and significantly so, when compared to the one with undetected births.\\

Consider an $n$-target hypothesis in FISST with $n-r$ births, out of which $s$ are detected, and the rest of the measurements are assigned to the target pdfs and false alarms according to some data association, the weight computation gives us:
\begin{align}
\omega_{FISST}^{q^{(r)}\sib_{n-r}\sn_a} = \eta (1-p_D)^{n-(m-k)}p_D^{(m-k)}\nonumber\\
\prod_{i=1}^{r}\int p(z_{\sigma^{(n)}_{a,i}}|x)p^{q(r),i-}(x)dx_i \nonumber\\
(\lambda_B)^{(n-r)}\bar{V}^{(n-r)-s}(\lambda_C)^k\omega_{\qr}.\label{SUB}
\end{align}
Let $s = (n-r)-1$. Now, consider an associated $(n-1)$-target hypothesis which has $(n-r)-1$ hypothesized births that are identical to the detected births in the $n$-target hypothesis above. Moreover, the underlying target pdfs, and data associations to clutter, births and targets is identical to the one above. Then, the weight computation becomes:
\begin{align}
\omega_{FISST}^{q^{(r)}\sib_{n-r-1}\sigma^{n-1}_a} = \eta (1-p_D)^{n-1-(m-k)}p_D^{(m-k)}\nonumber\\
\prod_{i=1}^{r}\int p(z_{\sigma^{(n-1)}_{a,i}}|x)p^{q(r),i-}(x)dx_i \nonumber\\
(\lambda_B)^{(n-r-1)}\bar{V}^{(n-r-1)- s}(\lambda_C)^k\omega_{\qr}.\label{0UB}
\end{align}
To simplify notation, let us denote the two hypotheses above as $v$ and $v'$ respectively. Then, along with the assumption $s = (n-r)-1$, it follows from equations \ref{SUB} and \ref{0UB} that:
\begin{equation}
\frac{\omega^v}{\omega^{v'}} = (1-p_D)(\lambda_B \bar{V}) = (1-p_D)\alpha \ll 1,
\end{equation}
since $p_D \approx 1$ and $\alpha \ll 1$. This shows that the weight of the hypothesis with all births detected, $v'$, is $\frac{1}{(1-p_D)\alpha}$ more likely than that with one birth undetected, $v$. \\
\begin{figure}[hbt]
	\centering
	\includegraphics[width=1.0\linewidth]{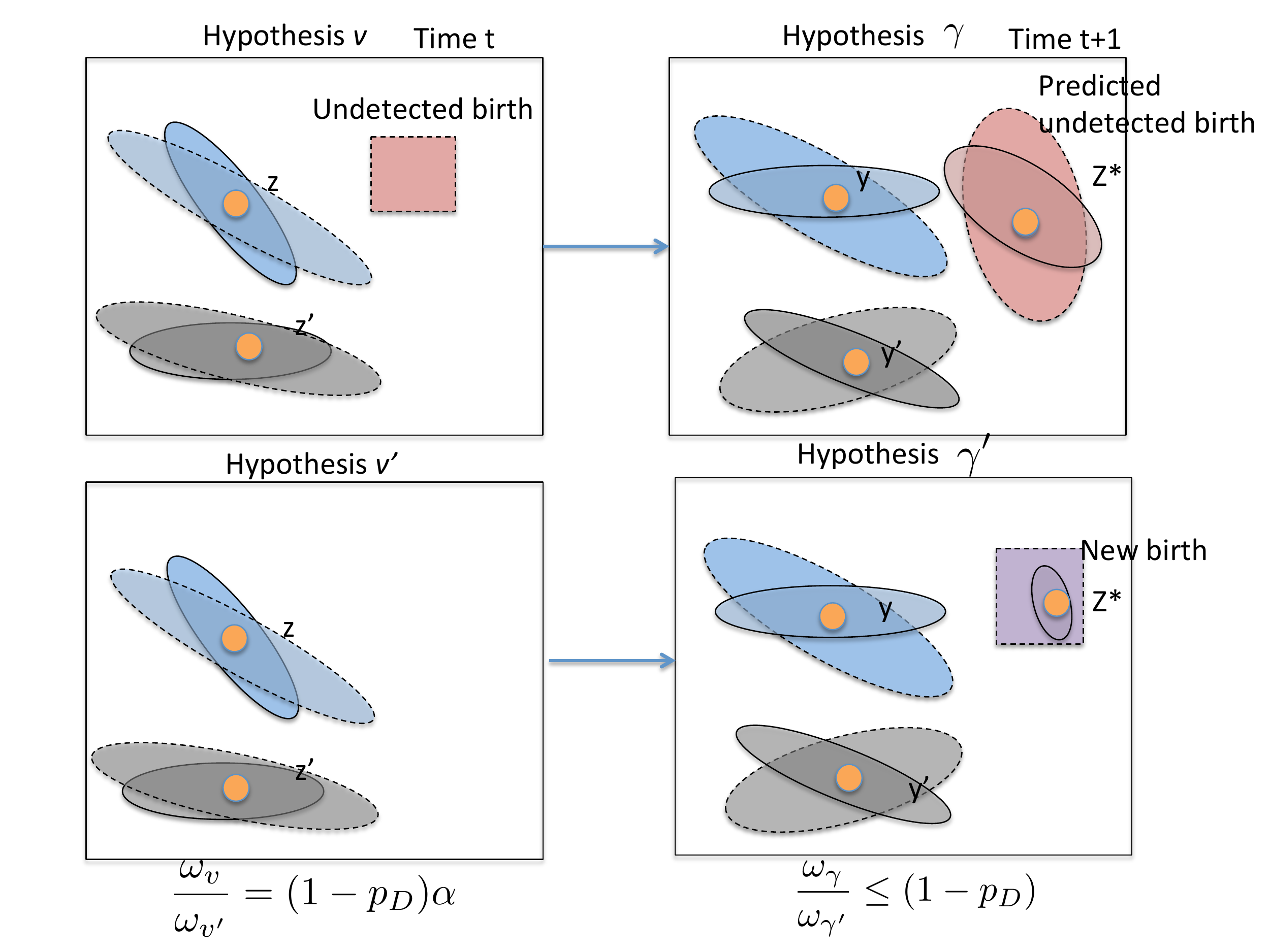}
	\caption{For any hypothesis with an undetected birth ($v$), there is an equivalent hypothesis with no undetected birth ($v'$) that has higher weight at time $t$, the only difference between the hypotheses is the undetected birth, and they are children of the same parent hypothesis. Moreover, for a child hypothesis of $v$ that has the undetected birth associated with $z^*$ (hypothesis $\gamma$), there is a child hypothesis of $v'$ that has one new birth which is associated with $z^*$ (hypothesis $\gamma'$), that still has higher weight. The dotted ellipses represent the predicted pdfs of the targets while the solid ones represent the updated pdfs, the observations are the orange dots.}
	\label{UB-construct}
\end{figure}

\begin{figure}[hbt]
	\centering
	\includegraphics[width=1.1\linewidth]{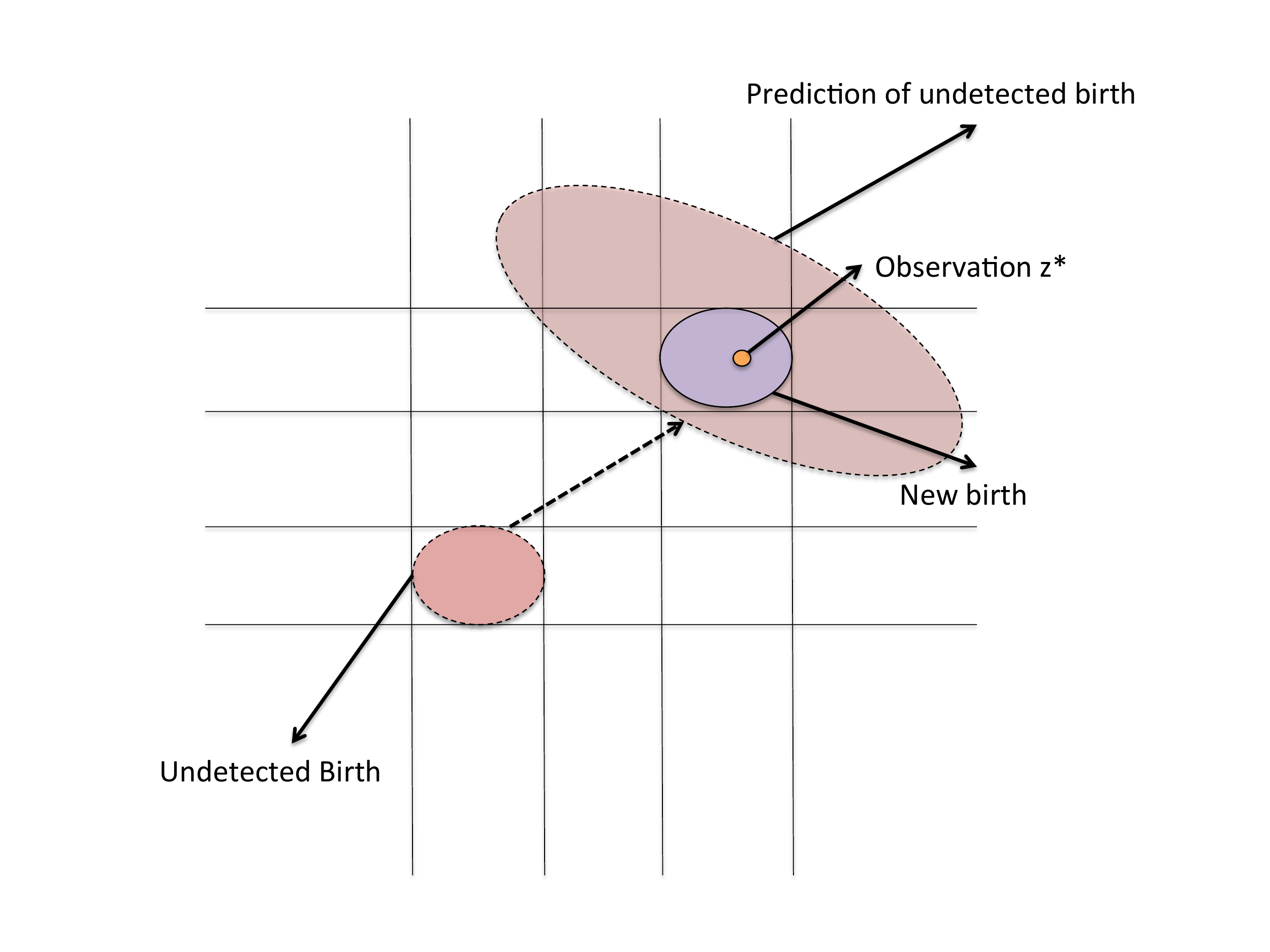}
	\caption{Prediction of an undetected birth: the uncertainty in the prediction of an undetected birth is always higher than that of a suitable new birth pdf making it more likely to associate an observation $z^*$ to the new birth than to the undetected birth.}
	\label{U-Birth}
\end{figure}

\noindent However, it may be that after prediction and update for the next time step, a suitable child hypothesis for $v$ might have more weight than a child hypothesis of $v'$. In the following, we will show that this is not possible under the typical assumptions in a tracking problem.\\
Reconsider the hypothesis $v$ and $v'$ from above and consider the following particular prediction and update step for them. The component pdfs underlying the two hypotheses are exactly the same except that $v$ has one extra undetected birth pdf. Now consider the following birth hypotheses for the two: $v$ has no birth and $v'$ has exactly one birth. Next consider the following data associations, the measurements associated to the targets/ clutter are exactly the same, except one measurement (say $z^*$) is assigned to the undetected birth for $v$, and the same measurement is assigned to a new birth in the case of $v'$. Let us call these children hypotheses $\gamma$ and $\gamma'$ respectively. Then:
\begin{equation}
\omega^{\gamma} = \bar{\eta} (1-p_D)^{n-(m-k)}p_D^{m-k} (A)(\int p(z^*/x)p_b^-(x)dx)\;\lambda_C^k \;\omega_v,
\end{equation}
where $p_b^-(.)$ represents the prediction of the undetected birth pdf, 
and similarly:
\begin{equation}
\omega^{\gamma'} = \bar{\eta} (1-p_D)^{n-(m-k)}p_D^{m-k}(A)\; \lambda_B \;\lambda_C^k \;\omega_{v'},
\end{equation}
where $A$ represents the product of the likelihoods arising from the detected existing targets common to both hypotheses. Thus,
\begin{equation}
\frac{\omega^{\gamma}}{\omega^{\gamma'}} = \frac{\int p(z^*/x)p_b^-(x)dx}{\lambda_B} \frac{\omega^v}{\omega^{v'}}.
\end{equation}
Consider the likelihood from the predicted birth pdf $\int p(z^*/x)p_b^-(x)dx$. We want to compare it to the likelihood of the observation coming from a new birth pdf in the pixel corresponding to $z^*$, say $b^*(.)$ (see Fig. \ref{U-Birth}). To simplify things, let the one step transition density be determined by the linear system $x'= Fx + Gw$, where $F$ and $G$ are suitable matrices, and $w$ is a white noise term. If the initial condition is a Gaussian with covariance $P$, the predicted covariance of the state is given by $P^-=FPF' + GQG'$. In general, due to the prediction step, $P^- > P$, i.e., the uncertainty in the state increases during the prediction. Suppose that all birth pdfs can be approximated by a suitable Gaussian pdf centered on the birth pixel, with covariance $P$, then predicted covariance of the birth pdf at the next step is given by $P^-$ as above. Therefore, the predicted covariance is strictly greater than the birth pdf's covariance. Consequently, the likelihood of the observation $z^*$ originating from the new birth is strictly higher than that of it originating from the predicted birth, i.e., we have that $\int p(z^*/x)p_b^-(x)dx < \frac{1}{\bar{V}}$. This situation is illustrated in Fig. \ref{U-Birth}.

Therefore, using this fact we have that:
\begin{equation}
\frac{\omega^{\gamma}}{\omega^{\gamma'}} < (1-p_D). \label{U0Comp}
\end{equation}
Thus, the development above shows that the hypothesis $\gamma'$ with a new birth is at least $\frac{1}{(1-p_D)}$ times more likely than the corresponding hypothesis $\gamma$ with an undetected birth. The analysis above is illustrated in Fig. \ref{UB-construct}.\\
In fact, with reference to Fig. \ref{U-Birth}, one can see that the inequality \ref{U0Comp} is conservative, and in reality, the likelihood of the observation coming from a new brith can be significantly more than that of it coming from an undetected birth, say $\int p(z^*/x)p_b^-(x)dx \leq \frac{1}{C\bar{V}}$, where $C>>1$. \\

A similar analysis to the one above can be used to show that given any hypothesis with $S$ undetected births, there is always a hypothesis with no undetected births whose weight is at least $\frac{1}{(1-p_D)^S}$ times the weight of the hypothesis with undetected births. In essence, the analysis above shows that it is much more difficult to detect an undetected birth as opposed to detecting a new birth. This seems logical since an undetected birth implies missing a birth target, a high $p_D$ value makes this difficult, which is then followed by detecting it at the next time step, which is unlikely, since it is easier to detect a suitable newly born target, since it has a lower uncertainty, at the next time step. 

\subsection{Discussion}
Thus far in this section, we have seen that the hypothesis weight update equations in FISST and HOMHT are identical for the case where there are a fixed number of targets. However, it does turn out that the HOMHT and FISST updates are indeed different in the case when there is target birth. In particular, we see that there is no provision in the HOMHT weight update law for undetected births, and thus, any HOMHT hypothesis detects every birth hypothesized within it. Also, due to the fact that the HOMHT treats birth similar to clutter, it results in a birth rate of $\lambda_B/p_D$ in HOMHT versus $\lambda_B$ for FISST. \\

Nonetheless, in our opinion, these discrepancies can be satisfactorily reconciled, at least in a practical sense. First, the birth Poisson process is only a model, and all models are wrong but the Poisson model happens to be useful. Thus, the birth parameter $\lambda_B$ should be treated as a tunable design parameter, and therefore the actual value of $\lambda_B$ will depend on the application at hand. Hence, if we treat the birth rate as a design parameter, the two methods can arrive at the same result with slightly differing values of $\lambda_B$. This is akin to the Extended Kalman filter where the designer tunes the value of the process noise covariance to suit the application at hand.\\

Second, and more importantly, we believe that the MHT ``heuristic" of assuming no undetected births is a practical way of approximating the combinatorially growing set of FISST hypotheses. Since, as shown above, there is always a hypotheses with no undetected birth that has significantly higher weight than one with undetected birth, at least $\frac{1}{1-p_D}$ times more, it follows that these are more probable. Hence, if our goal is to keep the top several hypotheses, then it follows that we need not consider ones with undetected birth. However, in such a case, the hypotheses weights, in general, would be off from the truth, but since the weights of the discarded hypotheses are much smaller when compared to the retained ones, this error should be negligible if the detection probability $p_D$ is sufficiently high.\\

Finally, albeit the mutli-target tracking problem is a hybrid problem (discrete hypothesis weights and continuous MT-pdfs), the structure of the problem dictates that the crux resides in the discrete hypothesis space, and thus, any MTT technique has to satisfactorily address the computational issue of tractable hypothesis management in order to be practically viable.

\section{Conclusion}
In this paper, we have presented a belief space perspective on the hypothesis dependent structure of the FISST recursions for multi-target tracking. We have also shown a unification of the hitherto deemed different FISST and MHT methodologies for multi-target tracking. It turns out that the two methods are identical for MTT problems with a fixed number of targets while they are different in the case of MTT problems with birth. Nonetheless, the  MHT heuristic of always detecting a new birth makes the problem of hypothesis management computationally tractable while being theoretically justifiable in that for every hypothesis with undetected births, there is always a corresponding hypothesis with no undetected birth that has significantly more weight. We have also developed a randomized approximation to the FISST approximation called the Randomized FISST (RFISST) that we have presented in a companion paper. Currently we are looking to apply the RFISST method to real data and develop large scale implementations that can scale to realistic Space Situational Awareness (SSA) scenarios. In other future work, we will look into guarantees regarding approximations to the FISST recursions such as the RFISST technique.

\appendix

\subsection{Proof of Proposition 3}
The proof of Proposition \ref{Prop2} is a special case of the proof of Proposition \ref{Prop3} with $\qn =1$, and hence, we only present the proof of Proposition \ref{Prop3} here. In the following, we drop the explicit reference to subscript $t$ for time, to declutter notation. 
\begin{proof}
Let the FISST predicted MT-pdf be given by:
\begin{align}
P(X,n) = \sum_{\qn} \omega^{\qn}p^{\qn}(X).
\end{align}
The MT-pdf in FISST is written as follows owing to the permutation of the arguments under the set representation:
\begin{align}
p^{\qn}(X) = \sum_{\nu} \prod_i  p^{\qn, \nu_i}(x_i),
\end{align}
where the $\nu$ are all possible permutations of the indices $\{1,2,\cdots n\}$, and $p^{\qn, i}(.)$ represents the $i^{th}$ component of the product in the MT-pdf. The FISST Bayesian update for the pdf given an MT-measurement $Z$  is:
\begin{align}
p(X,n/Z) = \frac{p(Z/X,n)\sum_{\qn}\omega^{\qn}p^{\qn}(X)}{\sum_k \frac{1}{k!} \int p(Z/X,k) \sum_{\rk}\omega^{\rk}p^{\rk}(X')dX'},
\end{align}
where the $k!$ factor is due to the interpretation of the denominator as a FISST set integral. In the above expression, let us concentrate on the term:
\begin{align}
T_{\qn} (X) = p(Z/x,n) \omega^{\qn} p^{\qn}(X) \nonumber\\
= \omega^{\qn} \sum_{\sn}p(\sn / n) \sum_{\nu} p(Z/\sn, X,n)  \prod_i p^{\qn, \nu_i}(x_i), \label{Eq0}
\end{align}
where $\sn$ represents any $n$-target data associations possible given the measurement $Z$. It may be seen that:
\begin{align}
p(X,n/Z) = \frac{\sum_{\qn}T_{\qn}(X)}{\sum_k \frac{1}{k!} \sum_{\rk}\int T_{\rk}(X)dX}.
\end{align}
Let 
\begin{align}
\eta_{\qn} \equiv  \frac{1}{n!} \int T_{\qn}(X) dX. \label{Eq1}
\end{align}
 Let $\sn$ be a particular data association that assigns exactly $k$ of the measurements to clutter, then:
\begin{align}
p(Z/\sn, X,n) = \frac{k!}{V^k} \prod_i p(z_{\sni}/x_i). \label{Eq2}
\end{align}
Then given any other data association $\bar{\sn}$ that assigns exactly $k$ measurements to clutter and a permutation $\nu'$, there always exists a unique k-clutter data association $\sn$ and an associated permutation $\nu$ such that:
\begin{align}
\prod_i p(z_{\bar{\sni}}/x_i)p^{\qn , \nu_i'} (x_i) = \prod_i p(z_{\sni}/x_{\nu_i})p^{\qn , i} (x_{\nu_i}). \label{Eq3}
\end{align}
Therefore, using \ref{Eq2}, \ref{Eq3} and \ref{Eq0} in \ref{Eq1}:
\small{\begin{align}
\eta_{\qn} \nonumber\\
= \frac{\omega^{\qn}}{n!}\sum_{\bar{\sn}} p(\bar{\sn}/n) \sum_{\nu'}\frac{k!}{V^k} \prod_i \int p(z_{\bar{\sni}}/x_i)p^{\qn,\nu_i'}(x_i) dx_i \nonumber\\
= \frac{\omega^{\qn}}{n!}\sum_{\sn} p(\sn/n)\sum_{\nu} \underbrace{\frac{k!}{V^k} \prod_i \int p(z_{\sni}/ x_{\nu_i}) p^{\qn, i} (x_{\nu_i}) dx_{\nu_i}}_{=\, l_{\qn\sn}\, \forall \, \nu}\nonumber\\
= \sum_{\sn} \omega^{\qn} p(\sn/n) l_{\qn\sn}. \nonumber
\end{align}}
The last line of the equation above follows from the fact that there are $n!$ permutations $\nu$. \\
Furthermore:
\begin{align}
T_{\qn}(X) \nonumber\\
= \omega^{\qn}\sum_{\sn} p(\sn/n)\sum_{\nu} \frac{k!}{V^k} \prod_i p(z_{\sni}/ x_{\nu_i}) p^{\qn, i} (x_{\nu_i}) \nonumber\\
= \omega^{\qn} \sum_{\sn} p(\sn/n) l_{\qn\sn} \sum_{\nu} \prod_i p^{\qn, \sn, i}(x_{\nu_i}),\nonumber\\
p^{\qn, \sn, i} (x) = \frac{p(z_{\sni}/x)p^{\qn,i}(x)}{\int p(z_{\sni}/x')p^{\qn,i}(x') dx'}.
\end{align}
where $p^{\qn,\sn, i}(.)$ is simply the $i^{th}$ pdf under $\qn$, $p^{\qn, i}(.)$ updated by the measurement $z_{\sn_i}$. Thus, it follows that:
\begin{align}
p(X,n/Z) \nonumber\\
= \sum_{\qn,\sn} p^{\qn, \sn} (X) \frac{\omega^{\qn} p(\sn/n)l_{\qn\sn}}{\sum_k \sum_{\rk} \sum_{\delta^{(k)}} \omega^{\rk} p(\delta^{(k)}/k) l_{\rk, \delta^{(k)}}}, \nonumber\\
p^{\qn, \sn} (X) = \sum_{\nu} \prod_i p^{\qn, \sn, i}(x_{\nu_i}).
\end{align}
\end{proof}

\subsection{Proof of Proposition 4}\label{Appendix:PredictionProof}
We provide a proof for Proposition 4 here, the proofs of Proposition 1 and the ``prediction" part of Proposition 3 are special cases of the same.
\begin{proof}
Let $X' = \{x_1',...,x_{n}\}$ be an $n$ component MT-state and let $X = \{x_1,\cdots x_r\}$ be an $r$ component MT-state. Then, it is clear that:
\begin{align} 
p(X'/ X, \sib_{n-r}) = \sum_{\nu}\prod_{i=1}^r p(x'_{\nu_i}/x_i)\prod_{i=r+1}^{n} p_{i-r}^{\sib}(x'_{\nu_i}),
\end{align}
where $\sib_{n-r}$ is any $n-r$ birth hypothesis. 
The primary task in proving Prop. 4 is to show Eq. \ref{P-B} for any birth hypothesis $\sib_{n-r}$. In the following, to simplify notation, we drop the explicit reference to $n-r$ in the birth hypothesis. 
\begin{align}
p_{\sib}^{\qr,-} (X')= \nonumber\\
\frac{1}{r!} \int \sum_{\nu} \prod_{i=1}^r p(x'_{\nu_i}/x_i)\prod_{i=r+1}^{n} p_{i-r}^{\sib}(x'_{\nu_i}) \nonumber\\
\sum_{\mu} \prod_{i=1}^r p^{\qr,\mu_i}(x_i)dx_i \nonumber\\
= \frac{1}{r!} \sum_{\nu, \mu} \prod_{i=1}^{n-r} p^{\sib}_i(x'_{\nu_{r+i}}) \prod_{i=1}^r \int p(x'_{\nu_i}/x_i)p^{\qr,\mu_i}(x_i)dx_i \nonumber\\
= \sum_{\nu_{r+1} ..\nu_n} \prod_{i=1}^{n-r} p^{\sib}_i(x_{\nu_{r+i}}) \nonumber\\
\frac{1}{r!} \sum_{\mu; \nu (r)}  \int \prod_{i=1}^r p(x'_{\nu_i}/x_i)p^{\qr,\mu_i}(x_i)dx_i,\label{P1}
\end{align}
where $\nu(r) = \{\nu_1,\nu_2,\cdots \nu_r\}$ represents the first $r$ elements of any n -permutation $\nu$. Given any $\nu(r)$ and $\bar{\nu}(r)$, there always exists a unique r-permutation $\bar{\mu}$ such that:
\begin{align}
\int  \prod_{i=1}^r p(x'_{\bar{\nu}_i}/x_i)p^{\qr, \bar{\mu}_i}(x_i) dx_i = \prod_{i=1}^r p^{\qr,i-}(x'_{\nu_i}). \label{P2}
\end{align}
Using Eq. \ref{P2} to simplify the second sum on the last line of Eq. \ref{P1},
\begin{align}
\sum_{\mu; \nu (r)}  \int \prod_{i=1}^r p(x'_{\nu_i}/x_i)p^{\qr, \mu_i}(x_i)dx_i = r! \sum_{\nu(r)} \prod_{i=1}^r p^{\qr,i-}(x'_{\nu_i}).
\end{align}
Hence, it follows that
\begin{align}
p^{\qr -}_{\sib_{n-r}}(X') 
= \sum_{\nu} \prod_{i=1}^r p^{\qr, i-}(x'_{\nu_i}) \prod_{i=r+1}^{n} p^{\sib_{n-r}}_{i-r}(x'_{\nu_i}),\nonumber\\
\end{align}
thereby proving the result.
\end{proof}

\subsection{Almost Sure Uniqueness of Tracks}
In this section, we shall show that the tracks are almost surely unique. In particular, we assume that the single target transition and likelihoods admit densities, and therefore, the probability of a particular observation emanating from a prior pdf is zero. We only show the result for the case of no target birth, its extension to the other case is reasonably straightforward.
\begin{proof}
In the following, we first show the reasoning under no clutter or missed detections.\\
Let $p_t^{(k)}(\omega)$ denote the pdf of the $k^{th}$ initial pdf/ target at time $t$, under the sample path $\omega \in \Omega$, with associated generated observations $z^{(k)}_{\tau}(\omega)$, for $\tau = 1\cdots t$. Recall that track $(k,l)$ corresponds to the pdf of the $k^{th}$ initial pdf under the observation sequence ${z_{\tau}^{(k,l)}}$, $\tau = 1\cdots t$. Note that the hypothesized observations $z^{(k,l)}_{\tau}$ are not the same as the actual observations emanating from the $k^{th}$ target, $z_{\tau}^{(k)}(\omega)$, since these observation associations are unknown.  Thus, the tracks result from associating any of the actual observations to any of the initial targets. \\
Let all the tracks at time $t-1$ be distinct. Consider the track $(k,l)$ at time $t$ with the end point $p_t^{(k,l)}$ (note that this is a sufficient statistic/ belief state for the entire history). Consider some other track $(m,n)$ with end point $p^{(m,n)}_t$. Then, it follows that $p_t^{(m,n)} = \tau(p_{t-1}^{(m,n')}, z^{(m,n)}_t)$ for some unique $n'$, where $\tau(p,z)$ represents the posterior pdf under observation $z$ given the prior pdf is $p$, and follows from the application of the prediction and update (with $z$) of the prior pdf $p$. In other words, the track end point $p_t^{(m,n)}$ is the child of some unique track end point $p_{t-1}^{(m,n')}$ from the previous time step, under the data association $z_t^{(m,n)}$. In order for the tracks to be indistinguishable, we need that $p_t^{(k,l)} = p_t^{(m,n)} =  \tau(p_{t-1}^{(m,n')}, z^{(m,n)}_t)$. Given $p_t^{(k,l)}$, due to the uniqueness of the the $\tau(.,.)$ map, and the distinctness of the tracks from the previous time step,  there can only be a unique $z^{(m,n)}_t$ (if any), say $z^*$, such that the above relationship is satisfied. However, the probability that any of the initial objects generates the observation $z^*$ is zero owing to the fact that the single target transition and likelihood probability measures admit densities, and noting that there is only a finite number of the objects, it follows that the probability that $p_t^{(m,n)} = p_t^{(k,l)}$ is zero. Noting that this is true for any selection of tracks $(k,l)$ and $(m,n)$, it follows that the tracks are almost surely distinct from one another at time $t$ if they are distinct at time $t-1$. We have already assumed that the initial pdfs  $p_0^{(k)}$ are distinct, therefore, it follows that the tracks are almost surely distinct for all times $t$.\\

In the presence of clutter and missed detections, the tracks can have data associations to observations due to clutter, as well as the null ($\phi$) observation. However, this merely increases the number of possible tracks (still finite), nonetheless, the argument above goes through by noting that the probability of getting the particular observation $z^*$ required for track indistinguishability, from Poisson clutter, is zero as well. This completes the proof of our assertion.
\end{proof}

\subsection{Incorporating Target Death} 
\begin{figure}[hbt]
	\centering
	\includegraphics[width=.9\linewidth]{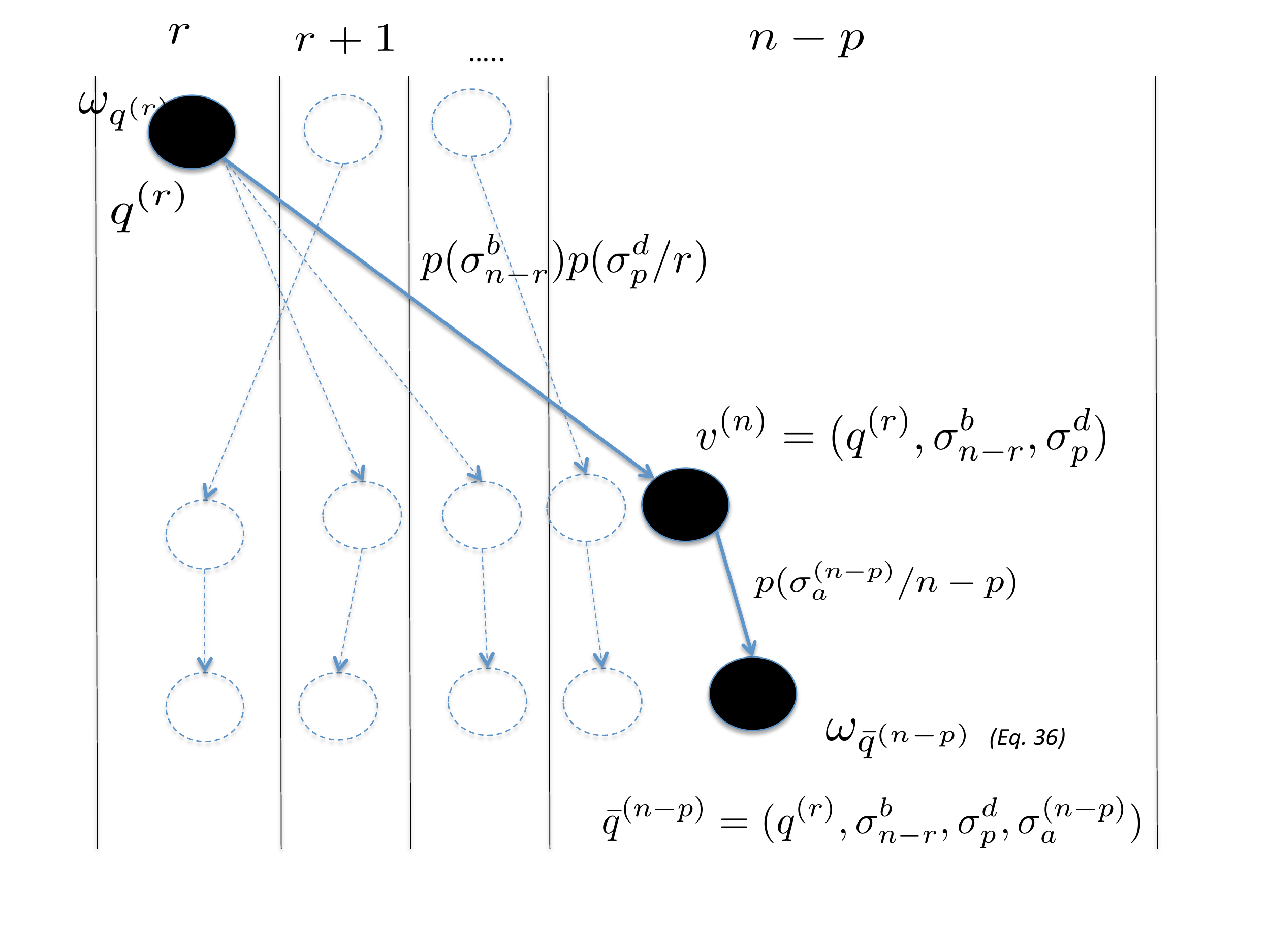}
	\caption{A schematic of the splitting of the hypothesis due to birth/ death of targets and data associations. Underlying each blob is a continuous multi-target pdf. A particular child and grandchild of a parent hypothesis, along with the transition probabilities, is outlined in bold, pictorially representing Eq. \ref{D-FISST}. }
	\label{HFISST}
\end{figure}

For the sake of completeness, we briefly include the case of target death in the following.
Given an $r$-target hypothesis $\qr$, the probability that a target survives can be modeled as a binomial process with the probability that a target survives equal to $\beta$, independent of any other target. Thus, the probability of the hypothesis that exactly $p$ of the targets out of $r$ survive, say $p(\sd_p|r)=\beta^p(1-\beta)^{r-p}$. Even with target death, the multi-target tracking problem remains that of tracking all possible descendants of the parent hypotheses, and thus, the primary difference with Eq. \ref{D-FISST} for the hypothesis update is that the survival probability $p(\sd_p|r)$ is now multiplied to the right of Eq. \ref{D-FISST*}, for all possible combinations of birth, survival and data association hypothesis:
\begin{align}
\omega_{\qr\sib_{n-r}\sd_p\sigma^{(n-p)}_a}\nonumber\\
= \eta l_{\qr\sib_{n-r}\sd_p\sigma^{(n-p)}_a} p(\sigma^{(n-p)}_a|n-p)p(\sd_p)p(\sib_{n-r})\omega_{\qr},  \label{D-FISST}
\end{align}
where note that the number of targets due to the $p$ deaths reduces to $n-p$, and therefore the data association hypotheses are for $n-p$ targets. The MT-pdf underlying the hypothesis is simply that of the birth hypothesis minus the pdfs of the targets that do not survive according to the hypothesis $\sd_p$. Note change in notation from Section \ref{sec:withBandD} such that: $\sd_p= \sigma^s_p$ and $\beta = p_s$.
This is also illustrated in Fig. \ref{HFISST}.

\bibliographystyle{IEEEtran}
\bibliography{dissertationReferences.bib}


\end{document}